%% file: submission-IT-2-CL-CLSR.tex
\documentclass[11pt, draftcls, onecolumn]{IEEEtran}

\def\withcolors{0}
\def\withnotes{0}

\input{packages}
\input{preamble}

\newcommand{\postrevision}[1]{#1}

\begin{document}
\title{Inference under Information Constraints II: Communication Constraints and Shared Randomness}
\date{}

\author{
  \IEEEauthorblockN{Jayadev Acharya\IEEEauthorrefmark{1}} 
  \and \IEEEauthorblockN{Cl\'{e}ment L. Canonne\IEEEauthorrefmark{2}}
  \and \IEEEauthorblockN{Himanshu Tyagi\IEEEauthorrefmark{3}}
}

\maketitle 

{
\renewcommand{\thefootnote}{}
  \footnotetext{%
      \IEEEauthorblockA{\IEEEauthorrefmark{1}Cornell University. Email: acharya@cornell.edu}\\
      \indent\IEEEauthorblockA{\IEEEauthorrefmark{2}IBM Research. Email: ccanonne@cs.columbia.edu}\\
      \indent\IEEEauthorblockA{\IEEEauthorrefmark{3}The Department of Electrical Communication Engineering,
        Indian Institute of Science, Bangalore 560012, India. Email: htyagi@iisc.ac.in}

      Jayadev Acharya is supported in part by the grant NSF-CCF-1846300 (CAREER), NSF-CCF-1815893, and a Google Faculty Fellowship. Part of this work was performed while Cl\'{e}ment Canonne was supported
 by a Motwani Postdoctoral Fellowship at Stanford University. Himanshu Tyagi is supported in part by a research grant from the Robert Bosch Center for Cyberphysical Systems (RBCCPS), Indian Institute of Science, Bangalore.   
}
\renewcommand{\thefootnote}{\arabic{footnote}}
\setcounter{footnote}{0}

\begin{abstract}
  \input{abstract}
\end{abstract}
\newpage

\thispagestyle{empty}
\setcounter{page}{1}

\tableofcontents

\ifnum\withnotes=1
  \clearpage
  \listoftodos
\fi

\newpage

\input{sec-introduction}

\input{sec-preliminaries}

\input{sec-setup}
\input{sec-simulation}

\input{sec-simulation-inference}

\input{sec-identity}

\section*{Acknowledgments}
The authors would like to thank the organizers of the 2018 Information
Theory and Applications Workshop (ITA), where the collaboration
leading to this work started.

\appendix
\input{app-simulation-impossibility}
\input{app-contracting-hashing}

\input{app-randomnessefficient-identity}
\input{app-identityfromuniformity}

\clearpage
  \bibliographystyle{IEEEtranS}
  \bibliography{references} 

\end{document}

%% file: packages.tex
\usepackage[T1]{fontenc}
\usepackage[utf8]{inputenc}
\usepackage{bbm}
\usepackage{xspace}                                     %
\usepackage{amsfonts,amsmath,amssymb, amsthm, mathtools}
\usepackage{thm-restate}
\usepackage{dsfont} %

\usepackage{algorithmicx,algpseudocode,algorithm}

\usepackage[usenames,dvipsnames,table]{xcolor}

\usepackage{multirow}
\usepackage{mfirstuc}

\usepackage{tikz}
\usetikzlibrary{calc,plotmarks}
\usetikzlibrary{arrows}

\usepackage[backref,colorlinks,citecolor=blue,bookmarks=true,linktocpage]{hyperref}
\usepackage{aliascnt}
\usepackage[numbered]{bookmark}
\usepackage[capitalise]{cleveref}

\usepackage[shortlabels]{enumitem}
\ifnum\withnotes=1
  \usepackage[colorinlistoftodos,textsize=scriptsize]{todonotes}
\fi

\usepackage{mleftright} %

\makeatletter
\let\pgfmathrandomX=\pgfmathrandom@
\usepackage{pgfplots}
\let\pgfmathrandom@=\pgfmathrandomX
\makeatother

\usepackage{tikz,forest}
\usetikzlibrary{arrows.meta}
\PassOptionsToPackage{hyphens}{url}

%% file: abstract.tex
A central server needs to perform statistical inference based on samples that are distributed over multiple users 
who can each send a message of limited length to the center.  We study  problems of distribution learning and identity testing
in this distributed inference setting and examine the role of shared randomness as a resource. We propose a general-purpose
\emph{simulate-and-infer} strategy that  uses only private-coin communication protocols and is sample-optimal for distribution learning. This general strategy turns
out to be sample-optimal even for distribution testing among private-coin protocols. Interestingly, we propose a public-coin
protocol that outperforms simulate-and-infer for distribution testing and is, in fact, sample-optimal. Underlying our public-coin protocol
is a random hash that when applied to the samples minimally contracts the chi-squared distance of their distribution to the uniform distribution.

%% file: sec-introduction.tex
\section{Introduction}
Sample-optimal statistical inference has come to the forefront of modern data analytics, where the sample size can be comparable
to the dimensionality of the data. In many emerging applications,
especially those arising in sensor networks and the Internet
of Things (IoT), we are not only constrained in the number of samples
but, also, are given access to only limited communication about the
samples. Similar concerns arise in federated learning where we want to analyze data distributed
across various users while requiring a limited amount of communication from each user.
We consider such a distributed inference setting and seek
sample-optimal algorithms for inference under communication
constraints.

In our setting, there are $\ns$ players, each of which gets a sample generated
independently from an unknown $\ab$-ary distribution and can send only $\numbits$ bits about their
observed sample to a central referee using a simultaneous message
passing (SMP) protocol for communication. The referee uses 
communication from the players to accomplish an inference task
$\task$; see~\cref{sec:setup} for formal definitions and problem formulation. We seek to answer the following question:

\begin{center} \emph{What is the minimum number of players $\ns$
required by an SMP protocol that successfully accomplishes $\task$, as
a function of $\ab$, $\numbits$, and the 
relevant parameters of $\task$?}
\end{center}
Our first contribution is a general \emph{simulate-and-infer} strategy for inference
under communication constraints where we use the communication to
simulate samples from the unknown distribution at the
referee. To describe this strategy, we introduce a natural notion of
\emph{distributed simulation}: $\ns$ players each observing an independent
sample from an unknown $\ab$-ary distribution $\p$ can send
$\numbits$ bits each to a referee. A distributed simulation protocol
consists of an SMP protocol and a randomized decision map that enables
the referee to generate a sample from $\p$ using the communication
from the players. Clearly, when\footnote{We assume throughout that $\log$ is in base 2, and for ease of discussion assume in this introduction that $\log\ab$ is an integer.} $\numbits\geq \log \ab$ such a sample can be obtained by getting the sample of any one player. But what can be
done in the communication-starved regime of $\numbits< \log \ab$?

We first show that perfect simulation is impossible using any finite number of
players in the communication-starved regime. But perfect simulation is
not even required for our application. When we allow a small probability of declaring
failure, namely admit Las Vegas simulation schemes, we obtain a 
distributed simulation scheme that requires an optimal
$\bigO{\ab/2^\numbits}$ players to simulate $\ab$-ary distributions
using $\numbits$ bits of communication per player. Thus, our proposed
simulate-and-infer strategy can accomplish $\task$ with a factor
$\bigO{\ab/2^\numbits}$ blow-up in
sample complexity.

The specific inference tasks we focus on are those of \emph{distribution learning}, where we seek to estimate the unknown
$\ab$-ary distribution to an accuracy of $\dst$ in total variation
distance, and \emph{identity testing} where we seek to know if the
unknown distribution is a pre-specified reference distribution $\q$ or at total variation
distance at least $\dst$ from it. For distribution learning, the simulate-and-infer strategy
matches the lower bound from~\cite{HOW:18} and is therefore
sample-optimal. For identity testing, the plot thickens.

Recently, a lower bound for the sample complexity of identity testing
using only private-coin protocols was established~\cite{ACT:18}. The simulate-and-infer
protocol is indeed a private-coin protocol, and we show that it achieves this lower 
bound. When public coins (shared randomness) are 
available,~\cite{ACT:18} derived a different, more relaxed lower
bound. The performance of simulate-and-infer is far from this lower
bound. Our second contribution is a public-coin protocol for identity testing
that not only outperforms simulate-and-infer but matches the lower
bound in~\cite{ACT:18} and is sample-optimal. 

We provide a concrete
description of our results in the next section, 
followed by an overview of our proof techniques in the subsequent
section. To put our results in context, we provide a brief overview of the
literature as well. 

\subsection{Main results}
We begin by summarizing our distributed simulation
results.\footnote{For simplicity of exposition, in the next result we 
allow the use of Las Vegas algorithms, which use variable number of players and produce a sample from
the unknown distribution when it terminates.
Equivalently, one may enforce a strict number of players
but allow the protocol to abort with a special symbol with small
constant probability, which is how our results will be stated
in~\cref{ssec:sampling:possibility}.} \vspace{-0.5em}  
\begin{theorem}\label{theo:distributed:simulation:informal}
  For every $\ab, \numbits\geq 1$, there exists a private-coin
  protocol with $\numbits$ bits of communication per player for
  distributed simulation over $[\ab]$ and expected number of players
  $\bigO{({\ab}/{2^\numbits})\vee 1}$. Moreover, this expected number is
  optimal, up to constant factors, even when public-coin and
  interactive communication protocols are allowed.
\end{theorem}
The proposed protocol only provides a relaxed guarantee, as the number
of players it requires is bounded only in expectation. In fact, we can show 
that distributed simulation is impossible, unless we allow for such
relaxation. 
\begin{theorem}\label{theo:distributed:simulation:impossiblity}
  For $\ab \geq 1$, $\numbits< \clg{\log\ab}$, and any $N\in\N$, there exist
  no SMP protocol with $N$ players and $\numbits$ bits of
  communication per player for distributed simulation over
  $[\ab]$. Furthermore, the result continues to hold even for
  public-coin and interactive communication protocols.
\end{theorem}
The proof is given in~\cref{ssec:sampling:impossibility}.

Since the distributed simulation protocol
in~\cref{theo:distributed:simulation:informal} is a private-coin protocol, we can use it to generate the desired number of samples from
the unknown distribution at the center to obtain the following
result.
\begin{theorem}[Informal]\label{theo:inference:simulation:informal}
  For any inference task $\task$ over $\ab$-ary distributions with
sample complexity $s$ in the non-distributed model, there exists a
private-coin protocol for $\task$ using $\numbits$ bits of
communication per player and requiring $\ns=O(s\cdot (\ab/2^\numbits\vee 1))$
players.
\end{theorem}
\noindent \postrevision{We note that the $O(\cdot)$ notation only hides absolute constants, and that the dependence on the inference task $\task$ is captured in the centralized sample complexity $s$.} Instantiating this general statement for 
 distribution learning and identity testing leads to the
following results.
\begin{corollary}\label{theo:uniformity:private:learning}
For every $\ab,\numbits\geq 1$, simulate-and-infer can accomplish
distribution learning over $[\ab]$, with $\numbits$ bits of
communication per player and
$\ns=\bigO{\frac{\ab^{2}}{(2^\numbits\wedge \ab)\dst^2}}$ players.
\end{corollary}
\begin{corollary}\label{theo:uniformity:private:randomness}
For every $\ab,\numbits\geq 1$, simulate-and-infer
can accomplish identity testing over $[\ab]$  using $\numbits$ bits
of communication per player and
$\ns=\bigO{\frac{\ab^{3/2}}{(2^\numbits\wedge \ab)\dst^2}}$ players. 
\end{corollary}
By the lower bound for sample complexity of distribution learning in~\cite{HOW:18}
(see, also,~\cite{ACT:18}), we note that simulate-and-infer is
sample-optimal for distribution learning even when public-coin
protocols are allowed. In fact, the sample complexity of
simulate-and-infer for identity testing matches the lower bound for
private-coin protocols in~\cite{ACT:18}, rendering it sample-optimal. 

Perhaps the most striking result in this paper is the next one, which shows that
public-coin 
protocols can outperform the sample complexity of private-coin
protocols for identity testing 
by a factor of $\sqrt{\ab/2^{\numbits}}$. 
\begin{theorem}\label{theo:uniformity:shared:randomness}
For every $\ab,\numbits\geq 1$, there exists a public-coin protocol 
for identity testing over $[\ab]$  using $\numbits$ bits
of communication per player and 
$\ns=\bigO{\frac{\ab}{(2^{\numbits/2}\wedge \sqrt{\ab})\dst^2}}$
players. 
\end{theorem}
\noindent \postrevision{Once again, this matches the lower bound for
public-coin protocols of~\cite{ACT:18}, showing our protocol is sample-optimal.} We further note that our protocol is quite simple to
describe and implement: We generate a  random partition of $[\ab]$
into $2^\numbits$ equisized parts and report which part each sample lies
in. Although, as stated, our protocol seems to require
$\Omega(\numbits\cdot \ab)$ bits of shared randomness, inspection of the proof shows that $4$-wise independent shared
randomness suffice, drastically reducing the number of random bits
required. See~\cref{rk:randomness:used} for a discussion. 

Our results are summarized in the table below. 
\renewcommand{\arraystretch}{1.25}
\begin{table}[H]
\centering
\caption{\label{table:results} Summary of the sample complexity of distributed learning and testing, under private and public randomness, for $\ab\geq 2^\numbits$. All results are order-optimal.}
\begin{tabular}{|>{\columncolor{red!10}} c|>{\columncolor{blue!10}} c|>{\columncolor{red!10}} c|>{\columncolor{blue!10}} c| }
\hline
  \multicolumn{2}{|c|}{Distribution Learning}
  & \multicolumn{2}{|c|}{Identity Testing} \\\hline 
   Public-Coin &
  Private-Coin & Public-Coin & Private-Coin \\\hline
  \hline 
   \multicolumn{2}{|c|}{\cellcolor{blue!50!red!10}$\frac{\ab}{\dst^2}\cdot \frac{\ab}{2^\numbits}$}
  & $\frac{\sqrt{\ab}}{\dst^2}\cdot \sqrt{\frac{\ab}{2^{\numbits}}}$ &
  $\frac{\sqrt{\ab}}{\dst^2} \cdot \frac{\ab}{2^{\numbits}}$ \\\hline
\end{tabular}
\end{table}
\subsection{Proof techniques}
We now provide a high-level description of the proofs of our main
results.

\paragraph{Distributed simulation}
The upper bound of~\cref{theo:inference:simulation:informal} uses a rejection-sampling-based approach; see~\cref{ssec:sampling:possibility}
for details. The lower bound follows by relating distributed
simulation to communication-constrained distribution learning and using the lower bound for
sample complexity of the latter from~\cite{HOW:18, ACT:18}.

\paragraph{Distributed identity testing}
For the ease of exposition, we hereafter focus on uniformity testing,
as it contains most of the ideas.  
To test  whether an unknown distribution $\p$ is uniform using at most
$\numbits$ bits to describe each sample, a natural idea is to randomly
partition the alphabet into $L\eqdef 2^\numbits$ parts, and send to
the referee independent samples from the $L$-ary distribution $\p'$
induced by $\p$ on this partition. For a random balanced partition
(\ie, where every part has cardinality $\ab/L$), clearly the uniform
distribution $\uniform_\ab$ is mapped to the uniform
distribution $\uniform_L$. Thus, one can hope to reduce the problem of
testing uniformity of $\p$ (over $[\ab]$) to that of testing
uniformity of $\p'$ (over $[L]$). The latter task would be easy to
perform, as every player can simulate one sample from $\p'$ and 
communicate it fully to the referee with $\log L = \numbits$ bits of
communication. Hence, the key issue is to argue that this random
``flattening'' of $\p$ would somehow preserve the distance to
uniformity. Namely, that if $\p$ is $\dst$-far from $\uniform_\ab$,
then (with a constant probability over the choice of the random
partition) $\p'$ will remain $\dst'$-far from $\uniform_L$, for some
$\dst'$ depending on $\dst$, $L$, and $\ab$. If true, then it is easy
to see that this would imply a very simple protocol with
$O(\sqrt{L}/{\dst'}^2)$ players, where all agree on a random
partition and send the induced samples to the referee, who then
runs a centralized uniformity test. Therefore, in order to apply the
aforementioned natural recipe, it suffices to derive a ``random
flattening'' structural result for $\dst' \asymp \sqrt{(L/\ab)}\dst$. 
 
An issue with this approach, unfortunately, is that the total variation
distance (that is, the $\lp[1]$ distance) does not behave as desired under
these random flattenings, and the validity of our desired result
remains unclear. 
Interestingly, an analogous statement with respect to the $\lp[2]$ distance turns out to
be much more manageable and suffices for our purposes. Specifically,
we show that a random flattening of $\p$ does preserve, with 
constant probability, the $\lp[2]$ distance to uniformity. In our
case, by the Cauchy--Schwarz inequality the original $\lp[2]$ distance
will be at least  
$\gamma\asymp\dst/\sqrt{\ab}$, which implies using known $\lp[2]$
testing results that one can test uniformity of the ``randomly
flattened'' $\p'$ with
$O(1/(\sqrt{L}\gamma^2))=O(\ab/(2^{\numbits/2}\dst^2))$ samples. This
yields the desired guarantees on the protocol.

\subsection{Related prior work}
The distribution learning problem is a finite-dimensional parametric learning problem, and the identity testing problem is a specific 
goodness-of-fit problem. Both these problems have a long history in statistics. However, the sample-optimal setting of interest to us has received a lot of attention in the past decade, especially in the computer science literature;  see~\cite{Rubinfeld:12:Survey,Canonne:15:Survey,BW:17}  for surveys.
 Most pertinent to our work is uniformity testing~\cite{GRexp:00,Paninski:08, DGPP:17}, the prototypical distribution testing problem for which the sample complexity was established to be $\Theta(\sqrt{\ab}/\dst^2)$ in~\cite{Paninski:08, VV:14}; as well as identity testing, shown to have order-wise similar sample complexity~\cite{BFFKRW:01,ADK:15,VV:14,DK:16,Goldreich:16}.

Distributed hypothesis testing and estimation
problems were first studied in information theory, although in a
different setting than what we consider~\cite{AhlCsi86, Han87,
HanAmari98}. The focus in that line of work has been to characterize
the trade-off between asymptotic error exponent and communication rate
per sample. 

Closer to our work is distributed parameter estimation and functional
estimation that has gained significant attention in recent years (see
\eg,~\cite{DJW:13,GMN:14,BGMNW:16,Watson:18}). In these works, much
like our setting, independent samples are distributed across players,
which deviates from the information theory setting described above
where each player observes a fixed dimension of each independent
sample. However, the communication model in these results differs from
ours, and the communication-starved regime we consider has not been
studied in these works.

The problem of distributed density estimation, too, has gathered
recent interest in various statistical
settings~\cite{BPCPE:11,BBFM:12,ZDJW:13,Shamir:14,DGLNOS:17, SzabovZ18, ZhuL18, HOW:18,XR:18,ASZ:18}. Among
these, our work is closest to the results in~\cite{HOW:18,HMOW-ISIT:18}
and~\cite{DGLNOS:17}. In  particular,~\cite{DGLNOS:17} considers both $\lp[1]$ 
(total variation) and $\lp[2]$ losses, although in a different setting
than ours. They study an interactive model where the
players do not have any individual communication constraint, but
instead the goal is to bound the {total} number of bits communicated
over the course of the protocol. This difference in the model leads to 
incomparable results and techniques (for instance, the lower bound for
learning $\ab$-ary distributions in our model is higher than the upper
bound in theirs).

Our current work further deviates from this prior literature, since we
consider distribution testing as well and examine the role of
public-coin for SMP protocols. Additionally, a central theme here is the
connection to distribution simulation and its limitation in enabling
distributed testing. In contrast, the prior work on distribution
estimation, in essence, establishes the optimality of simple protocols
that rely on distributed simulation for inference. We note that
although recent work of~\cite{BCG:17} considers both communication
complexity and distribution testing, their goal and results are very
different~--~indeed, they explain how to leverage on negative results
in the standard SMP model of communication complexity to obtain sample
complexity lower bounds in {collocated} distribution testing.

Problems related to joint
simulation of probability distributions have been the object of focus
in the information theory and computer science literature. 
Starting with the works of G{\'a}cs and K{\"o}rner~\cite{GK:73} and
Wyner~\cite{Wyner:75} where the problem of generating shared
randomness from correlated randomness and vice-versa, respectively,
were considered, several important variants have been studied such as
{correlated sampling}~\cite{Broder:97,KT:02,Holenstein:07,BGHKRS:16}
and {non-interactive simulation}~\cite{KA:12,GKS:16,DMN:18}.
Yet, our problem of {exact} simulation of a single (unknown)
distribution with communication constraints from multiple parties has
not been studied previously to the best of our knowledge.

\subsection{Relation to chi-square contraction lower bounds}
This work is the second of a series of papers, the first of
which (\cite{ACT:18}) presented a general technique for
establishing lower bounds for inference under information
constraints. When information constraints are imposed, the statistical
distances shrink due to the data processing inequality. 
At a high-level, the lower bound in~\cite{ACT:18} was
based on quantifying the contraction in chi-square distance in a
neighborhood of the uniform distribution due to information
constraints. Note that in view of the reduction
in~\cref{app:identity:from:uniformity}, the neighborhood of any
distribution is roughly isometric to the neighborhood of the
uniform distribution (though the isometry can depend on the reference
distribution). Thus, our lower bound aptly captures the bottleneck
imposed by information constraints for a broad class of inference
problems, and not just uniformity testing. 

The current article, and our
upcoming article~\cite{ACFT:19},\footnote{See~\cite{ACFT-AISTATS:19} for a
preliminary version.}{} seeks to find schemes that match the lower
bounds established in~\cite{ACT:18}. An interesting feature of our
lower bounds is that they quantitatively differentiate the chi-square
contraction caused by private- and public-coin
protocols. Our schemes in this paper draw on the principles
established by our lower bounds in~\cite{ACT:18} and use a \emph{minimally contracting
hash} for inference under information constraints. Specifically, 
our private-coin simulate-and-infer scheme and public-coin scheme
are based on identifying a private-coin and public-coin communication
protocol, respectively, that minimally contract the chi-square
distances in the neighborhood of the uniform distribution. We term this
principle of designing inference schemes under information constraints
 the \emph{minimally contracting hashing} (MCH) principle. At this
point, it is just a heuristic where we seek mappings that attain the
minmax and maxmin chi-square contractions that appear in our lower
bounds in~\cite{ACT:18}, and propose them as a good candidate for
selecting channels for inference under information constraints in our
setting. We believe, however, that a formal version of the MCH principle can
be established and applied gainfully in this setting.

The MCH principle seems to remain valid even for local privacy constraints, as 
considered in~\cite{ACFT:19}. Moreover, in addition to the papers in this
series, our preliminary calculations suggest that our treatment and the MCH principle extend to testing
problems concerning high-dimensional distributions as well. Finally, while in
this paper we have quantified the reduction in sample complexity due
to availability of public randomness for
a fixed amount of communication per sample, quantifying the complete
sample-randomness tradeoff for distributed identity testing under
communication constraints is work in progress.

\subsection{Organization}
We begin by formally introducing our
distributed model in~\cref{sec:setup}. Next,~\cref{sec:simulation}
introduces the question of distributed simulation and contains our
protocols and impossibility results for this problem. In~\cref{sec:simulate:infer}, we consider the 
relation between distributed simulation and private-coin
distribution inference. The subsequent section,~\cref{sec:identity}, focuses on the problem
of identity testing and contains the proof of~\cref{theo:uniformity:shared:randomness}.

%% file: sec-preliminaries.tex
\section{Notation and preliminaries}\label{sec:preliminaries}

Throughout this paper, we denote by  $\log$ the logarithm to the base
$2$. We use standard asymptotic
notation $\bigO{\cdot}$, $\bigOmega{\cdot}$, and $\bigTheta{\cdot}$ for complexity orders,\footnote{Namely, for two non-negative sequences $(a_n)_{n\in \N}$ and $(b_n)_{n\in \N}$, we write $a_n = O(b_n)$ (resp., $a_n = \Omega(b_n)$) if there exist $C>0$ and $N\geq 0$ such that $a_n \leq Cb_n$ (resp., $a_n \geq Cb_n$) for all $n\geq N$. Further, we write $a_n = \Theta(b_n)$ when both $a_n = O(b_n)$ and $a_n = \Omega(b_n)$ hold.} and, for two non-negative sequences, write
$a_n \lesssim b_n$ to indicate that there exists an absolute constant
$c>0$ such that $a_n \leq c\cdot b_n$ for all $n$. Finally, we will
denote by $a\wedge b$ and $a\vee b$ the minimum and maximum of two
numbers $a$ and $b$, respectively.

Let $[\ab]$ be the set of integers $\{1,2,\dots,\ab\}$. Given a fixed (and known)
discrete domain $\domain$ of cardinality $|\cX|=\ab$, we write
$\distribs{\ab}$ for the set of probability distributions over
$\domain$, \ie, \[ \distribs{\ab}
= \setOfSuchThat{ \p\colon[\ab]\to[0,1] }{ \normone{\p}=1 }\,.  \]
For a discrete set $\domain$, we denote by $\uniformOn{\domain}$ the
uniform distribution on $\domain$ and will omit the subscript when the
domain is clear from context. 

The \emph{total variation distance} between two probability
distributions $\p,\q\in\distribs{\ab}$ is defined
as \begin{equation*} \totalvardist{\p}{\q} \eqdef \sup_{S\subseteq\domain} \left(\p(S)-\q(S)\right)
= \frac{1}{2} \sum_{x\in\domain} \abs{\p(x)-\q(x)}, \end{equation*}
namely, $\totalvardist{\p}{\q}$ is equal to half of the $\lp[1]$ distance
of $\p$ and $\q$. In addition to total variation distance, we will
extensively use the $\lp[2]$ distance between distributions $\p,\q\in\distribs{\ab}$,
denoted $\normtwo{\p-\q}$.

%% file: sec-setup.tex
\section{The setup:  Communication, simulation, and inference protocols}\label{sec:setup}

\subsection{Communication protocols}

We restrict ourselves to \emph{simultaneous message passing} (SMP) protocols of
communication, wherein the messages from all players are transmitted
simultaneously to the central server, and no other communication is
allowed. We allow randomized SMP protocols and distinguish between two
forms of randomness: private-coin protocols, where each player can
only use their own independent private randomness that is not
available to the referee and public-coin protocols, where the players
and the referee have access to shared randomness. SMP rules out any
other interaction between the players except the agreement on the
protocol and coordination using shared randomness for public-coin SMP
protocols. In particular, this setting precludes interactive
communication models.  Nonetheless, this setting is natural for a
variety of use-cases where players represent users connected to a
central server or sensors connected to a fusion center. It can even be
used for the case where each sample is seen by the same machine, but
at different times, and the machine does not maintain any memory to
store the previous samples. For instance, this machine can be an
analog-to-digital converter that quantizes each input to $\numbits$
bits. \postrevision{Even in this noninteractive setting, we note that the use of shared randomness 
arises naturally in, \eg, asymmetric settings where the central server can broadcast sporadically a common 
random seed to the users; or when this random seed is hardcoded in the sensors 
before they are deployed.}

\begin{definition}[Private-coin SMP Protocols]\label{d:private-prot}     
Let $U_1, \dots, U_\ns$ denote independent random variables, which are also independent jointly of $(X_1, \dots, X_\ns)$, and represent the private randomness of the players. An $\numbits$-bit \emph{private-coin} SMP protocol $\pi$ consists of the following two steps: (a)~Player $i$ selects their channel\footnote{\postrevision{Following the convention in information theory, we define a channel $W$ from $\cX$ to $\cY$ as a randomized mapping $W\colon\cX\to \cY$. We represent it by a $|\cY|\times|\cX|$ \emph{transition probability matrix} $W$ whose rows and columns are indexed by $y\in \cY$ and $x\in\cX$, respectively, and its $(y,x)$th entry
$W(y \mid x)\eqdef W_{y,x}$ is the probability of observing $y$ when the input to
the channel is $x$.}}{} $W_i\in\cW_\numbits$ as a function of $U_i$, (b)~and sends their message $M_i\in\{0,1\}^\numbits$, which is obtained by passing $X_i$ through $W_i$, to the referee. The referee receives the messages $M=(M_1, M_2, \dots, M_\ns)$, but does not have access to the private randomness $(U_1,\dots, U_\ns)$ of the players. 
\end{definition}

We assume that the protocol is decided ahead of time, namely the distribution of $U_i$s is known to the referee, but not the realization. Note that in a private-coin SMP communication protocol, the communication $M_i$ from player $i$ is a randomized function of $(X_i, U_i)$.  Moreover, since both $(X_1, \dots, X_\ns)$ and $(U_1, \dots, U_\ns)$ are generated from a product distribution, so is $(M_1,\dots,M_\ns)$.

\begin{definition}[Public-coin SMP  Protocols]
Let $U$ be a random variable independent of $(X_1, \dots, X_\ns)$, available to all players and the referee. An $\numbits$-bit \emph{private-coin} SMP protocol $\pi$ consists of the following two steps: (a)~Players select their channels $W_1,\ldots, W_\ns\in\cW_\numbits$ as a function of $U$, and (b)~send their messages $M_1,\ldots, M_\ns\in\{0,1\}^\numbits$,  by passing $X_i$ through $W_i$, to the referee. The referee receives the messages $M=(M_1, \dots, M_\ns)$ and is given access to $U$ as well.
\end{definition}
 In contrast to private-coin protocols, in a public-coin SMP communication protocol, the communication $M_i$ from player $i$ is a (randomized) function of $(X_i, U)$ and therefore the $M_i$s are not independent. They are, however, independent conditioned on the shared randomness $U$.
 
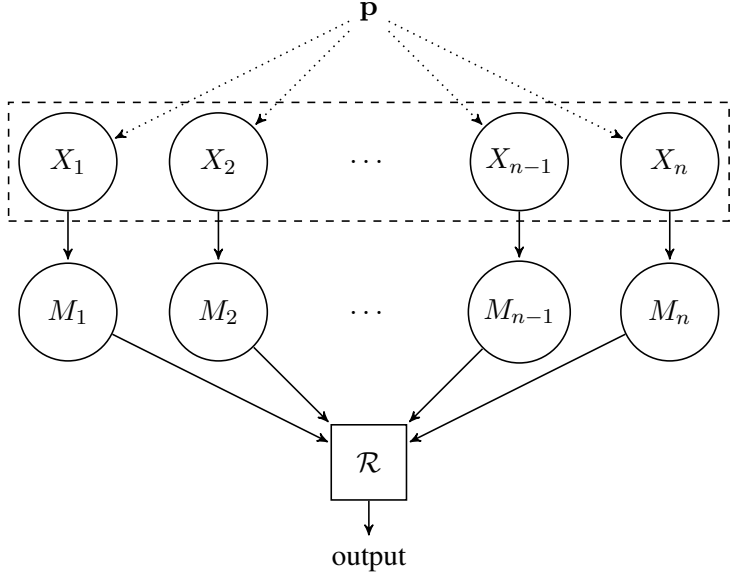
\begin{figure}[ht]\centering
\caption{\label{fig:model}The communication-constrained distributed model, where each $M_i\in\{0,1\}^\numbits$. In the private-coin setting the channels $M_1,\dots,M_\ns$ are independent, while in the public-coin setting they are jointly randomized.}
\input{fig-ic}
\end{figure}

 We denote the communication protocols that are used at the players to
 generate the messages by $\pi$.
For public-coin
   protocols, to make explicit the role of the randomness in the
   choice of the channels, we sometimes write $\pi(x^\ns,u)$ to denote
   the output of the protocol (messages) when the  
 input of the players is $x^\ns=(x_1, \dots, x_\ns)$ and the
 public-coin realization is $U=u$. Also, 
 we write $\pi_i(x^\ns, u)$ for the message sent by player $i$ using protocol $\pi$. See~\cref{fig:model} for a depiction of the communication setting.

\subsection{Distributed simulation protocols}\label{ssec:intro:simulation}
The distributed simulation problem we propose is rather natural, yet, to the best of our knowledge, has not been studied in prior literature. In this section, we will define the simulation problem, and in the next section exhibit its use as a natural tool to solve {any}
communication-limited inference problem. Recall that our goal is to
enable the referee to generate samples from the unknown distribution
using communication from the players. Note that players only know the
alphabet $[\ab]$ from which samples are generated, but have no other
knowledge of the distribution. We allow the players to use an SMP
protocol, private-coin or public-coin, to facilitate simulation of
samples by the referee.

We now state the question of simulation formally.  An $\numbits$-bit
\emph{simulation protocol} $\cS = (\pi, T)$ of $\ab$-ary distributions
using $\ns$ players consists of an $\numbits$-bit SMP protocol $\pi$
and a decision mapping $T$. The output of $\pi$ is an element in
$\cM^\ns$, where $\cM=\{0,1\}^\numbits$. The decision mapping
$T\colon\cM^\ns\to\cX\cup\{\bot\}$ is a randomized function that takes as
input the messages from the players and outputs an element in
$\cX\cup\{\bot\}$, \postrevision{where $\bot$ is the ``abort'' (no outcome) symbol}.  {Upon receiving messages $m^\ns=(m_1, \dots,
m_\ns)\in \cM^\ns$, the referee
outputs $x\in\cX$ with probability $\probaOf{T(m^\ns)=x}$ 
 and the symbol $\bot$ with probability $T(\bot \mid
 m^\ns)=1-\sum_{x\in \cX}\probaOf{T(m^\ns)=x}$}. {Interpreting the randomized function $T$ as a channel
with input alphabet $\cM^n$ and the output alphabet $\cX\cup \{\bot\}$,
we denote $\probaOf{T(m^\ns)=x}$ by $T(x\mid m^\ns)$.} The protocol is private-coin 
if $\pi$ is a private-coin communication protocol, and it is
public-coin if $\pi$ is public-coin. For public-coin protocols, the
decision mapping $T=T_U$ can be chosen as a function of $U$, the public
randomness. We want the distribution of the random output of the decision mapping to 
coincide with the unknown underlying distribution $\p$. This
objective is made precise next.   

\begin{definition}[$\alpha$-Simulation]
\label{def:simulation}
A protocol $S=(\pi, T)$ is an \emph{$\alpha$-simulation} protocol if for every $\p\in \distribs{\ab}$ that generates the input samples $X_1,
\dots, X_\ns$ for the SMP protocol $\pi$, the output $\hat{X}=T(\pi(X_1,\ldots,X_\ns))\in
\cX\bigcup\{\bot\}$ of the 
simulation protocol $T$ satisfies
\[
\probaDistrOf{X^\ns \sim \p^\ns}{\hat{X} =x \mid \hat{X} \neq \bot} =
\p_x, \quad \forall\, x\in \cX,
\]
and the probability of abort satisfies
\[
\probaDistrOf{X^\ns \sim \p^\ns}{\hat{X} =\bot} \leq \alpha.
\] 
A $0$-simulation, namely a simulation with probability of abort zero,
is termed \emph{perfect simulation}.
\end{definition}

\subsection{Distributed inference protocols}
We give a general, decision-theoretic description of distributed
inference protocols that is applicable beyond the use-cases considered
in this work. For the most part, we will restrict to learning and
identity testing of discrete distributions, but our results for
distributed inference are valid for general settings. 

We start with a description of inference tasks. An inference problem
$\task$ is a tuple $(\class,\cX, \cE, \loss)$, where $\class$ is a
collection of distributions over $\cX$, $\cE$ is a class of allowed
actions or decisions that can be taken upon observing samples
generated from $\p\in \class$, and $\loss\colon
\class\times\cE \to \R_+^q$ is a loss function used to evaluate the
performance. A (randomized) decision rule is a map $e\colon \cX^\ns\to
\cE$, and for samples $X^\ns$  generated from $\p\in\class$, the loss
of the decision rule is measured by the vector $\loss(\p, e(X^\ns))$ in $\R_+^q$. Our
benchmark for performance will be the expected loss vector 
\begin{align}
L(\p, e)\eqdef \bE{X^\ns\sim\p}{\loss(\p, {e}(X^\ns)}\,.
\label{eq:loss}
\end{align}
Note that the expected loss vector, too, is a $q$-dimensional vector.

An $\numbits$-bit \emph{distributed inference protocol} $\cI =
(\pi,e)$ for the inference problem $(\class, \cX, \cE, \loss)$ consists of
an $\numbits$-bit SMP protocol $\pi$ and an estimator $e$ available to
the referee who, upon observing the messages $M=(M_1,\ldots, M_\ns)$, and follows a (randomized) decision rule $e\colon\cM^\ns\to \cE$. For private-coin inference protocols, $\pi$ is a private-coin SMP protocol, and for  public-coin inference protocols, both the communication protocol $\pi$ and the decision rule $e$ are allowed to depend on the public randomness $U$, available to everyone. \postrevision{The expectation in~\eqref{eq:loss} is then taken over both $X^\ns$ and the randomness (private or public) of the protocol.}

We now state a measure of performance of inference
protocols.
\begin{definition}[$\vec{\gamma}$-Inference protocol]
\label{def:inference}
For $\vec{\gamma}\in\R_+^q$, a protocol $(\pi, e)$ is a
\emph{$\vec{\gamma}$-inference protocol} if, for every $\p\in\class$,
\[
L_i(\p, e) \leq \gamma_i, \quad\, \forall 1\leq i \leq
q\,,
\]
where $L_i(\p, e)$ denotes the $i$th coordinate of $L(\p, e)$.
\end{definition}
We instantiate the abstract definitions above with two illustrative
examples that we study in this paper.

\paragraph{Distribution Learning}  In the $(\ab, \dst)$-distribution
learning problem, we seek to estimate a distribution $\p$ in
$\distribs{\ab}$ to within $\dst$ in total variation
distance. Formally, a (randomized) mapping $e\colon\cX^\ns
\to\distribs{\ab}$ constitutes an $(\ns,\dst, \delta)$-estimator for
$\distribs{\ab}$ if the estimate $\hat 
\p=e(X^\ns)$ satisfies
\[
\sup_{\p\in\distribs{\ab}}\probaDistrOf{X^\ns\sim
  \p}{\dtv\Paren{\hat{\p},\p}>\dst}< \delta,
\]
where $\dtv(\p,\q)$ denotes the total variation distance between $\p$
and $\q$. Namely, $\hat {\p}$ estimates the input distribution $\p$ to
within distance $\dst$ with probability at least $1-\delta$.

The sample complexity of $(\ab, \dst, \delta)$-distribution learning
is the minimum $\ns$ such that there exists an
$(\ns,\dst,\delta)$-estimator for $\distribs{\ab}$. It is well-known that the
sample complexity of distribution learning is $\Theta(\ab/\dst^2)$ and
the empirical distribution  attains it.

This problem can be cast in our general framework by setting
$\cX=[\ab]$, $\class=\cE=\distribs{\ab}$, $q=1$, and
$\loss(\p, \hat{\p})$ is given by
\[
\loss(\p, \hat{\p}) \eqdef \indic{\totalvardist{\p}{\hat {\p}}> \dst}.
\]
For this setting of distribution learning, we term the $\delta$-inference protocol an
$\numbits$-bit $(\ab, \dst, \delta)$-\emph{learning protocol} for
$\ns$ players.

\paragraph{Identity Testing} Let  $\q\in \distribs{\ab}$ be a known reference distribution. In the $(\ab, \dst, \delta)$-identity testing
problem, we seek to use samples from unknown $\p\in\distribs{\ab}$ to test if $\p$ equals $\q$ or if it is $\dst$-far from $\q$ in total variation distance. Specifically, an $(\ns, \dst, \delta)$-test is given by a (randomized) mapping $\Tester\colon\cX^\ns\to\{0,1\}$ such that
\begin{align*}
\probaDistrOf{X^\ns\sim \p^\ns}{\Tester(X^\ns)=1}>1-\delta, &\text{ if
} \p=\q,\\ \probaDistrOf{X^\ns\sim \p^\ns}
             {{\Tester(X^\ns)=0}}>1-\delta, &\text{ if }
             \totalvardist{\p}{\q}>\dst.
\end{align*}
Namely, upon observing independent samples $X^\ns$, the algorithm
should ``accept'' with high constant probability if the samples come
from the reference distribution $\q$ and ``reject'' with high constant
probability if they come from a distribution significantly far from
$\q$.

The sample complexity of $(\ab, \dst,\delta)$-identity testing is the minimum $\ns$ for which an $(\ns,\dst,\delta)$-test exists for $\q$. While this quantity can depend on the reference distribution $\q$, it is customary to consider sample complexity over the worst-case
$\q$.\footnote{The sample complexity for a fixed $\q$ has been studied
  under the ``instance-optimal'' setting (see~\cite{VV:14,BCG:17}):
  while the question is not fully resolved, nearly-tight upper and
  lower bounds are known.} In this worst-case setting, while it has
been known for some time that the most stringent sample requirement
arises for $\q$ set to the uniform distribution, a recent result
of~\cite{Goldreich:16} provides a formal reduction of arbitrary $\q$
to the uniform distribution case. It is therefore enough to consider
$\q=\uniformOn{\ab}$, the uniform distribution over $[\ab]$; identity testing for $\uniformOn{\ab}$ is termed the $(\ab,\dst,\delta)$-\emph{uniformity testing} problem. For constant $\delta$, the sample complexity of
$(\ab,\dst)$-uniformity testing was shown to be
$\Theta\big(\sqrt{\ab}/\dst^2\big)$ in~\cite{Paninski:08,VV:14}, and the exact dependence on $\delta$ was later identified in~\cite{HuangM13,DGPP:17}.

Uniformity testing, too, can be obtained as a special case of our
general formulation by setting $\cX = [\ab]$,
$\class=\{\uniformOn{\ab}\}\cup\{\p\in\distribs{\ab}:\totalvardist{\p}{\uniformOn{\ab}}>\dst\}$,
$\cE = \{0,1\}$, and the $2$-dimensional loss function $\loss\colon
\class\times\cE\to\R^2$ to be \begin{align*}
  \loss_1(\p, b) &= b\cdot \indic{\p= \uniformOn{\ab}}\,,\\
  \loss_2(\p, b) &= (1-b)\cdot \indic{\p\neq \uniformOn{\ab}}\,,
\end{align*} for $b
\in \{0,1\}$. For simplicity, we consider the error parameter
$\vec{\gamma}=(\delta,\delta)$.\footnote{We observe that by this
  formulation allows, more generally, to study the dependence of
  sample complexity Type-I and Type-II error probabilities $\delta_1$
  and $\delta_2$ by considering $\vec{\gamma}=(\delta_1, \delta_2)$.}
For this case, we term the $\delta$-inference protocol an
$\numbits$-bit $(\ab, \dst, \delta)$-\emph{uniformity testing
  protocol} for $\ns$ players. We provide $(\ab,
\dst,\delta)$-uniformity testing protocols for arbitrary $\delta$, but
we establish lower bounds only for $\delta=1/12$.  This choice of
probability of error is to remain consistent with~\cite{ACT:18},
since we borrow the general lower bounds from there. For simplicity we
will refer to $(\ab, \dst, 1/12)$-uniformity testing protocols simply
as $(\ab, \dst)$-\emph{uniformity testing protocols}.

Note that distributed variants of several other inference problems
such as that of estimating functionals of distributions and parametric
estimation problems can be included as instantiations of the
distributed inference problem described above.

%% file: fig-ic.tex
\begin{tikzpicture}[->,>=stealth',shorten >=1pt,auto,node distance=20mm, semithick]
  \node[circle,draw,minimum size=13mm] (A)
  {$X_1$}; \node[circle,draw,minimum size=13mm] (B) [right of=A]
  {$X_2$}; \node (C) [right of=B] {$\dots$}; \node[circle,draw,minimum
  size=13mm] (D) [right of=C] {$X_{\ns-1}$}; \node[circle,draw,minimum
  size=13mm] (E) [right of=D] {$X_\ns$};

  \node[draw,dashed,fit=(A) (B) (C) (D) (E)] {};
  o \node[circle,draw,minimum size=13mm] (YA) [below of=A]
  {$M_1$}; \node[circle,draw,minimum size=13mm] (YB) [below of=B]
  {$M_2$}; \node (YC) [below of=C]
  {$\dots$}; \node[circle,draw,minimum size=13mm] (YD) [below of=D]
  {$M_{\ns-1}$}; \node[circle,draw,minimum size=13mm] (YE) [below
  of=E] {$M_\ns$};

  \node (P) [above of=C] {$\p$}; \node[rectangle,draw, minimum
  size=10mm] (R) [below of=YC] {$\referee$}; \node (out) [below
  of=R,node distance=13mm] {output};

  \draw[->] (P) edge[dotted] (A)(A) edge (YA)(YA) edge
  (R); \draw[->] (P) edge[dotted] (B)(B) edge (YB)(YB)
  edge (R); \draw[->] (P) edge[dotted] (D)(D)  edge
  (YD)(YD) edge (R); \draw[->] (P) edge[dotted] (E)(E) 
  edge (YE)(YE) edge (R); \draw[->] (R) edge (out);
\end{tikzpicture}

%% file: sec-simulation.tex
\section{Distributed simulation}\label{sec:simulation}
In this section, we consider the distributed simulation problem
described in~\cref{ssec:intro:simulation}. We start by considering the
more ambitious problem of perfect simulation, where using a finite number of players
$\ns$, the referee must simulate a sample from the unknown $\p$ using
the $\numbits$-bit messages from the players. We then consider the 
relaxed problem of $\alpha$-simulation for a constant
$\alpha\in(0,1)$ (see~\cref{def:simulation}). We prove the following
results for these problems. 
\begin{enumerate} 
	\item In~\cref{ssec:sampling:impossibility}, we show that for
          any $\numbits< \clg{\log \ab}$ and finite $\ns$, perfect simulation is impossible using
$\ns$ players. 

\item In~\cref{ssec:sampling:possibility}, for any constant
$\alpha\in(0,1)$, we exhibit an $\numbits$-bit private-coin
$\alpha$-simulation protocol for $\ab$-ary distributions using
$O((\ab/2^\numbits)\log(1/\alpha))$ players. 

\item {Finally, in~\cref{sec:sampling:optimality:simulation}, drawing
  on the lower bounds for distribution learning, we will prove the
sample-optimality of our distributed simulation algorithm above up to
constant factors. In fact,  even with public coins
  the number of players cannot be reduced by more than a constant
  factor.}
\end{enumerate}

We have defined the distributed simulation problem as one where the
output distribution conditioned on not outputting $\bot$ is identical
to $\p$. One may wonder about another natural relaxation to perfect
simulation, where the goal is to generate a sample according to a
distribution that is $\alpha$-close to $\p$ (say, in total variation
distance). A primary reason for considering the former is that the
ability to generate samples 
from $\p$ will allow us to compose it with a centralized algorithm for
any inference task, as we show
in~\cref{sec:simulate:infer}. 

\subsection{Impossibility of perfect simulation when $\numbits<\log \ab$}\label{ssec:sampling:impossibility} 
We show that any simulation that works for all points in the interior 
of the $(\ab-1)$-dimensional probability simplex must fail for a
distribution on the boundary.  Our main result of this section is the
following:
\begin{theorem}\label{theo:sampling:impossibility:non:adaptive:k:l}
For any $\ns\geq 1$, there exists no $\numbits$-bit perfect simulation
for $\ab$-ary distributions using $\ns$ players unless $\numbits \geq
\clg{\log \ab}$.
\end{theorem}
\begin{proof}
Suppose that for $\numbits<\clg{\log \ab}$ there exists an $\numbits$-bit
(public-coin) perfect simulation $\cS=(\pi, T)$ for $\ab$-ary
distributions using $\ns$ players. Fix a realization $U=u$ of the
public randomness.  Since $\numbits<\clg{\log \ab}$, by the pigeonhole
principle for each player at least two symbols in $[\ab]$ map to the
same message. Therefore, we can find a message vector $(m_1, \dots,
m_\ns)$ and distinct elements $x_i, x_i'\in [\ab]$ for each $i\in [n]$
such that
\begin{align}
\label{eqn:sim-impossibility}
\pi_i(x_i, u) = \pi_i(x_i',u) = m_i,
\end{align} 
that is for $U=u$, the SMP protocol sends the same message vector $m$
when the observation of players is $(x_1, \dots, x_\ns)$ or $(x_1', \dots,
x_\ns')$. For a perfect simulation, the referee is not allowed to
output $\bot$, and it must output a symbol in $[\ab]$.

Next, consider a message $m$ and a symbol $x\in[\ab]$ such that
$T_u(x\mid m)>0$, namely the referee outputs $x$ with a nonzero
probability when the public randomness is $U=u$ and the message received is
$m$. The key observation in our proof is that since $x_i\neq x_i'$ in
view of~\eqref{eqn:sim-impossibility}, for each $i$ either $x_i\neq x$
or $x_i^\prime \neq x$. Without loss of generality, we assume that
$x_i\neq x$ for each $1\leq i\leq \ns$.

Finally, consider a distribution $\p$ such that $\p_x=0$ and
$\p_{x'}>0$ for all $x'\ne x$. For perfect simulation, under this
distribution, the referee must never declare $x$. However, conditioned
on the public-coin realization being $U=u$, the probability of
observing the message $(m_1, \dots, m_\ns)$ above is
\[
{\probaCond{M=(m_1, \dots, m_\ns)}{U=u} =\sum_{\tilde{x}} \prod_{i=1}^\ns
W_{i,u}(m_i \mid \tilde{x}_i)p(\tilde{x}_i) \geq
\prod_{i=1}^nW_{i,u}(m_i \mid x_i)\p_{x_i} > 0,}
\]
where $W_{i,u}$ denotes the channel used by player $i$ to sent its message when the public randomness is $U=u$.
Thus, the referee has a nonzero probability of outputting
$x$ {given $U=u$}, even though $\p_x=0$. This {contradicts} the assumption that
$\cS$ is a perfect simulation. 
\end{proof}
Note that the proof above shows that any perfect simulation of a
distribution $\p$ in the interior of the $(\ab-1)$-dimensional
probability simplex must fail for at least one distribution on the
boundary of the simplex. In fact, a much stronger impossibility result
holds. For the smallest non-trivial parameter values of $\ab = 3$ and
$\numbits=1$, no perfect simulation protocol exists that simulates all
distributions in any open neighborhood in the interior of the
probability simplex.
\begin{theorem}
  \label{theo:sampling:impossibility:non:adaptive:k3}
For any $\ns\geq 1$, there does not exist any $\numbits$-bit perfect
simulation of ternary distributions ($\ab=3$) unless $\numbits\geq 2$,
even when the input distribution is known to {come} from an open
set in the interior of the probability simplex.
\end{theorem} 
We defer the proof of this theorem to~\cref{app:simulation:impossibility}. Roughly speaking, the argument
proceeds by establishing that we can, without loss of generality,
restrict to deterministic protocols. We then 
show that any deterministic simulation protocol must output $\bot$
with a nonzero probability -- contradicting the assumption of perfect
simulation. Together, the two incomparable impossibility results
of~\cref{theo:sampling:impossibility:non:adaptive:k:l,theo:sampling:impossibility:non:adaptive:k3}
(one for general $1\leq \numbits < \clg{\log \ab}$ but at the boundary of
the probability simplex; the other for {$\numbits=1$} and $\ab \geq 3$,
but in the interior) rule out perfect simulation in a strong sense in
the case of SMP protocols.

We close this section by extending our impossibility result to beyond
SMP protocols, to the setting where the players are allowed to communicate interactively.\footnote{Public-coin protocols do allow the players to coordinate using shared randomness. But they do not interact in any other way.}
In a \emph{\postrevision(sequentially) interactive communication protocol}, players $1$ to $\ns$
communicate sequentially in rounds, with  
player $i$ communicating in round $i$. The communication is in a
broadcast mode where, along with the referee, the players too receive
communication from each other. The communication of player $i$ can
depend on their local observation and 
the communication received in the previous $i-1$ rounds from the other
players. \postrevision{We hereafter omit the word ``sequentially,'' and simply refer to such protocols as \emph{interactive communication protocols}.}

Our next result shows that perfect simulation is impossible, even when
players use an interactive communication protocol. 
The proof uses a standard method  for simulating 
sequential protocols with SMP protocols, by increasing the number of
players (see, for instance, reduction of
round complexity in~\cite{KushilevitzNisan97}). 
\begin{lemma}
For every $\ns\geq 1$, if there exists an interactive
public-coin $\numbits$-bit perfect simulation of $\ab$-ary distributions with
$\ns$ players, then there exists a
public-coin $\numbits$-bit perfect simulation of $\ab$-ary
distributions with {$2^{\numbits\ns}$} players that uses only SMP.
\end{lemma}
\begin{proof}
Consider an interactive communication protocol $\pi$ for
distributed simulation with $\ns$ players and
$\numbits$ bits of communication per player.  We can view the overall
protocol as a $2^{\numbits}$-ary tree of depth $\ns$ where
each node is assigned to a player. An execution of the protocol is a path
from the root to the leaf of the tree, namely along any such path each player appears once. %
This protocol can be simulated non-interactively using at most
$ (2^{\numbits\ns}-1)/(2^\numbits-1) <
        2^{{\numbits\ns}}$ players, where players $(2^{j-1}+1)$ to $2^j$ send all messages
correspond to nodes at depth $j$ in the tree. Then, the
referee receiving all the messages can output the index of the leaf node by following
the path from root to the leaf.
\end{proof}
{In other words, any interactive protocol with a finite number of players can be simulated by a non-interactive (\ie, SMP) protocol with a finite (albeit exponentially larger) number of players.  As our impossibility results hold for non-interactive protocols with \emph{any} finite number of players, the above lemma therefore implies that they still hold for \emph{interactive} communication protocols.}
\begin{corollary}
\cref{theo:sampling:impossibility:non:adaptive:k:l,theo:sampling:impossibility:non:adaptive:k3} 
hold even when the players are allowed to use interactive
communication protocols for simulation.
\end{corollary}

\subsection{An $\alpha$-simulation protocol using rejection sampling}
\label{ssec:sampling:possibility}

In this section we present our construction of a simulation protocol
for $\ab$-ary distributions
using $\ns=O(\ab/2^\numbits)$ players, establishing the following theorem:
\begin{theorem}
  \label{theo:distributed:simulation}
For every $\alpha\in(0,1]$ and $\numbits\geq 1$, there exists an $\numbits$-bit $\alpha$-simulation
of $\ab$-ary distributions using 
\[
40 \clg{\log\frac{1}{\alpha}}\cdot\clg{\frac{\ab}{2^{\numbits}-1}}
\] players. Moreover, the protocol is deterministic for the players, and only requires private randomness at the referee.
\end{theorem}
At a high level, our algorithm divides players into batches and
constructs a $3/4$-simulation using each batch. The overall simulation
declares the output symbol of the first batch that does not declare an
abort. By using
$O(\clg{\log\frac{1}{\alpha}})$ batches, we can boost the probability of
abort from $3/4$ to $\alpha$. 

To simplify the presentation, we first present the protocol
for $\numbits=1$ and analyze its performance.
Even for this case, we build our protocol in steps, starting with the
basic version given in~\cref{alg:simulation-ell-1-basic} below, which
requires $\ns=2\ab$ players.
\begin{algorithm}[ht]
\begin{algorithmic}[1]
\Require $\ns=2\ab$ players observing one independent sample each from
an unknown $\p$
\State For $1\leq i\leq \ns$, players $(2i-1)$ and $2i$ send one bit to indicate whether their
observation is $i$.
\State The referee receives these $\ns=2\ab$ bits $M_1, \dots, M_\ns$.
\If{exactly one of the bits $M_1, M_3, \dots, M_{2\ab-1}$ is equal to one, say the bit $M_{2i-1}$,
 {and} the corresponding bit $M_{2i}$ is zero,}
{the referee outputs $\hat{X}= i$;}
\Else{ the referee outputs $\bot$ (abort).}
\EndIf
\end{algorithmic}
\caption{Distributed simulation protocol using $\numbits=1$: The basic version}
\label{alg:simulation-ell-1-basic}
\end{algorithm}
The next result characterizes the performance of this simulation
protocol.
\begin{theorem}
The protocol in~\cref{alg:simulation-ell-1-basic} uses $2\ab$ players
and is a $3/4$-simulation for $\p\in \distribs{\ab}$ such that
$\norminf{\p}\leq 1/2$. 
\end{theorem}
\begin{proof}
From the description of the protocol,  it is easy to verify that 
the output $\hat X$ of the protocol takes the value $i$ with probability
\begin{align}
\label{eq:prob-i}
\bPr{\hat X = i}=  \p_i\cdot \prod_{j\ne i} (1-\p_j) \cdot (1-\p_i)= \p_i\cdot \prod_{j=1}^{\ab} (1-\p_j)\,,
\end{align}
where the first term in the product corresponds to $M_{2i-1}$ being
$1$, the second term to all the other messages from odd-numbered
players being $0$, and
the final term for $M_{2i}$ to be $0$. Note that this probability is
proportional to $\p_i$, showing that conditioned on the event $\{\hat
X\in [\ab]\}$, the output is indeed distributed according to
$\p$. 

Next, we bound the probability of abort for this
protocol. By summing~\eqref{eq:prob-i} over all $i$ in $[\ab]$, we
obtain that the probability $\rho_{\p} \eqdef
\probaOf{\referee \text{ does not output } \bot}$ is  given by
\[
\rho_{\p}= \prod_{j=1}^{\ab} (1-\p_j).
\]
Observe that while (as discussed above), conditioned on success, the
output is from $\p$, the probability of abort can depend on $\p$. In
particular, if there is one symbol with large probability (close to
one), the success probability can be arbitrarily close to zero. This
is where we use our assumption $\norminf{\p}\leq 1/2$ to establish that
\[
\rho_{\p}= \prod_{j=1}^{\ab} (1-\p_j)\ge \frac14.
\label{lem:ub:prod}
\]
Indeed, the claimed bound follows from observing that $1-x\geq 1/4^x$ for all $x\in [0, 1/2]$.
Therefore, the probability of aborting is bounded above by $3/4$, completing the proof.
\end{proof}
To handle the case when $\norminf{\p}$ may exceed
$1/2$, we consider the distribution $\q$ on $[2\ab]$ defined by
\[
    \q_{i} =\q_{\ab+i}= \frac{1}{2}\cdot \p_i,\qquad i\in[\ab]\,.
\] 
This distribution satisfies the condition $\norminf{\q}\leq 1/2$, and
therefore, the previous protocol yields $3/4$-simulation for it using
$4\ab$ players observing independent samples from $\q$. The problem
now reduces to obtaining samples from $\q$ using samples from
$\p$, and then obtaining back a sample from $\p$ given a sample from $\q$ generated by the referee. Towards that, we note that although the players do not know $\p$,
given a sample 
from $\p$, it is 
easy to convert it into a sample from $\q$ as follows. Player $j$ upon receiving $X_j\sim\p$, maps it to $X_j$ or $X_j+\ab$ with equal
probability. We can use this process to convert samples from $4\ab$
players to sample from $\q$ and apply~\cref{alg:simulation-ell-1-basic}
to simulate a sample $\tilde X$ from $\q$ at the referee. Finally, 
we can convert the sample $\tilde X$ from $\q$ to that from $\p$ by
declaring {$\hat X = (\tilde{X}-1 \bmod \ab)+1$}. Our enhancement
of~\cref{alg:simulation-ell-1-basic} described next does exactly this,
with a slight modification to avoid the use of additional randomness
at the players (but instead using randomness at the referee only). 
\begin{algorithm}[ht]
\begin{algorithmic}[1]
\Require $\ns=4\ab$ players observing one independent sample each from
an unknown $\p$
\State Players divide themselves in two sets of $2\ab$ players each,
and each set executes a copy of~\cref{alg:simulation-ell-1-basic}.

\State The referee receives message bits $(M_1,\dots, M_{4\ab})$ from all
the players, and
independently flips each message bit that is $1$ to $0$ with
probability $1/2$ to obtain $(\overline{M}_1,\dots, \overline{M}_{4\ab})$. \label{step:coupling-q}

\If{exactly one of the message bits $\overline{M}_1, \overline{M}_3, \dots, \overline{M}_{4\ab-1}$
  is $1$, say the message $\overline{M}_{2i-1}$, {and} the corresponding message sequence $\overline{M}_{2i}$ is ${0}$,}
\If{$i>\ab$,}{ the referee updates $i\gets  i-\ab$;}
\EndIf
\State{ the referee outputs $\hat X = i$;}
\Else{ the referee outputs $\bot$.}
\EndIf
\end{algorithmic}
\caption{Distributed simulation protocol using $\numbits=1$: The enhanced version}
\label{alg:simulation-ell-1-enhanced}
\end{algorithm}
This protocol achieves our desired performance for the case $\numbits=1$.
\begin{theorem}\label{theo:generate:sample:1bit:1sample}
The protocol in~\cref{alg:simulation-ell-1-enhanced} uses $4\ab$ players
and is a $3/4$-simulation for $\p\in \distribs{\ab}$. Moreover, the communication
protocol used by the players is a deterministic protocol. 
\end{theorem}
\begin{proof}
We first establish the following claim.
\begin{claim}
The distribution of flipped bits obtained
after~\cref{step:coupling-q} coincides with that for
  message bits when we execute~\cref{alg:simulation-ell-1-basic} using
  samples from $\q$.
\end{claim}
To see this, note that, for
$i\in[\ab]$, players $i$ and $i+\ab$ send the message $1$ with
probability $\p_i$ each. Therefore, the flipped bits of these players
will equal $1$ with probabilities $\q_i=\p_i/2$ each. But this is
exactly the probability with which these messages would be $1$ if the
samples of the players were generated from $\q$ and we were
executing~\cref{alg:simulation-ell-1-basic}. 

Next, note that the operation of the referee from here on can be
described alternatively as obtaining $\tilde{X}$ by
executing~\cref{alg:simulation-ell-1-basic} for $2\cdot 2\ab=4\ab$ samples from
$\q$ and declaring {$\hat X = (\tilde{X}-1 \bmod \ab)+1$} if $\tilde{X}\neq
\bot$. Thus, the overall protocol behaves as if the players and the
referee executed~\cref{alg:simulation-ell-1-basic} for samples from
$\q$ and then the referee declared the output ${}\bmod \ab+1$, if it was
not a $\bot$. As we saw above, this protocol constitutes a
$3/4$-simulation for $\p$.
\end{proof}

Moving now to the more general setting of arbitrary $\numbits\in\{1,\ldots, \clg{\log \ab}\}$, we
simply modify~\cref{alg:simulation-ell-1-enhanced} to use the extra
bits of communication. For simplicity, we assume that $2^\numbits -1$
divides $\ab$ and set $m\eqdef\ab/(2^{\numbits}-1)$. We partition the domain $[\ab]$ into
$m$ equal contiguous parts $S_1, \ldots, S_m$,
with $|S_i|=2^\numbits-1$. Our proposed modification
to~\cref{alg:simulation-ell-1-enhanced} to extend it for $\numbits\geq 1$
is given in~\cref{alg:simulation-ell}.

\begin{algorithm}[ht]
\begin{algorithmic}[1]
\Require $\ns=4m$ players observing one independent sample each from
an unknown $\p$

\State Players $2j-1, 2j, 2(j+m)-1, 2(j+m)$, $1\leq
j \leq m$, send the following communication
depending on their observed sample $x$:

\If{$x\notin{S_j}$,}{ send the all zero sequence $\mathbf{0}$ of length $\numbits$.}
\Else{ indicate the precise value of $x\in S_j$ using the remaining
  $2^\numbits-1$ binary sequences of length $\numbits$. We denote the
  sequence sent for $i\in S_j$ by
  $s_i\in\{0,1\}^\numbits\setminus\{\mathbf{0}\}$.}
\EndIf

\State The referee independently changes the message $M_j$ from player
$j$ that is not $\mathbf{0}$ to $\mathbf{0}$ with probability $1/2$, to obtain the flipped
message $\overline{M}_j$.

\If{exactly one of the message sequences $\overline{M}_1, \overline{M}_3, \dots, \overline{M}_{4m-1}$
  is nonzero, say the message $\overline{M}_{2j-1}$, {and} the
  corresponding message sequence $\overline{M}_{2j}$ is ${\bf
    0}$,}\label{step:2m-compare} 
\If{$j>m$,}{ the referee updates $j\gets  j-m$;}
\EndIf
\State{ if $\overline{M}_{2j-1}=s_i$, the referee outputs $\hat X = i
  \in S_j$;}
\Else{ the referee outputs $\hat X= \bot$.}
\EndIf
\end{algorithmic}
\caption{Distributed simulation protocol using $\numbits\geq 1$:
  Basic block}
\label{alg:simulation-ell}
\end{algorithm}
The previous protocol can be developed incrementally in the same
manner as the protocol for $\numbits=1$. First, we obtain a protocol
under some additional assumption on $\p$ using
$2\clg{\frac{\ab}{2^{\numbits}-1}}$ players and then circumvent the
requirement for that assumption by converting samples from $\p$ into
samples for $\q$ by doubling the number of players. The form above is
obtained in the same manner as that
of~\cref{alg:simulation-ell-1-enhanced}, by relegating the requirement
for randomization at the players to the referee.  

The performance of this protocol is characterized in the theorem below. 
\begin{theorem}
  \label{theo:generate:sample:lbits:1sample}
  For any $\numbits\geq 1$,~\cref{alg:simulation-ell} uses
  $4\clg{\frac{\ab}{2^{\numbits}-1}}$ players and is a
  $3/4$-simulation for $\p\in \distribs{\ab}$. Moreover, the communication
protocol used by the players is a deterministic protocol.
\end{theorem}
\begin{proof}
The proof is similar to that
of~\cref{theo:generate:sample:1bit:1sample}, with appropriate
extensions to handle $\numbits>1$. Note that the players in the set
$\cP_j\eqdef \{2j-1, 2j,
2(j+m)-1,2(j+m)\}$, $j \in [m]$, use the same mapping to determine the
message to send. Let $i\in S_j$. Then, for all players in the set $\cP_j$,
the flipped message equals $s_i$ (the sequence representing message
$i$) with probability $\p_i/2$. It follows that the flipped message is
$\mathbf{0}$ for any of these players with probability
$(1-\p(S_j)/2)$.  Denoting $j_i$ the $j\in[m]$ such that $i\in S_j$, note
that only players in $\cP_{j_i}$ can declare $s_i$ with positive
probability. Therefore, by combining the previous observations with
the fact that the messages of all players are independent, we get
\[
\bPr{\hat X=i}= 2\cdot \frac{\p_i}{2} \cdot \prod_{j\ne j_i}\left(1-\frac{\p(S_j)}2\right)\cdot
    \left(1-\frac{\p(S_{j_i})}2\right),
\]
where the first factor of $2$ represents two cases where 
$\overline{M}_{2j_i-1}=s_i$  or $\overline{M}_{2(j_i+m)-1}=s_i$,
$\prod_{j\ne j_i}(1-\p(S_j)/2)$ is the probability that each
of the flipped messages $\overline{M}_{2t-1}$  is $\mathbf{0}$ for
$t\neq j_i$ or $t\neq j_i+m$, and the final factor $(1-\p(S_{j_i}/2))$ is the
probability that $M_{2t}=0$ for $t=j_i$ or $t=j_i+m$. As a
consequence, we get that 
\[
\bPr{\hat X\neq\bot} = \prod_{j\in [m]}\left(1-\frac{\p(S_j)}2\right)\geq \frac 1 4,
\]
where in the final bound we used once again the fact that $1-x\geq
1/{4^x}$ for $0\leq x\leq 1/2$. This completes the proof. 
 \end{proof}
Finally, we boost the probability of successful simulation from $1/4$
to {$1-\alpha$} by using multiple blocks. 

\begin{algorithm}[ht]
\begin{algorithmic}[1]
\Require $\ns = 40 \clg{\log\frac{1}{\alpha}}\cdot\clg{\frac{\ab}{2^{\numbits}-1}}$
players observing one independent sample each from an unknown $\p$

\State Divide players into $10\clg{\log\frac{1}{\alpha}}$ disjoint groups of
$4\clg{\frac{\ab}{2^{\numbits}-1}}$ players each.

\State Execute~\cref{alg:simulation-ell} to each block successively,
one block at a time. 

\If{ all blocks do not declare $\bot$ as the output,}{ output $\hat
  X=i$ where $i\in[\ab]$ is the output of the first block that does not output $\bot$;} 
\Else{ output $\hat X= \bot$ and terminate.}
\EndIf
\end{algorithmic}
\caption{Distributed simulation protocol using $\numbits\geq 1$:
  Complete protocol}
\label{alg:simulation-ell-complete}
\end{algorithm}
We conclude with {the} proof establishing
that~\cref{alg:simulation-ell-complete} attains the performance
claimed in~\cref{theo:distributed:simulation}.
\begin{proofof}{\cref{theo:distributed:simulation}}
Each group in~\cref{alg:simulation-ell-complete}
executes the $3/4$-simulation protocol given
in~\cref{alg:simulation-ell}, and the overall protocol outputs the symbol in
$[\ab]$ that the first group to succeed outputs, if such a group
exists. This is a simple rejection sampling procedure, and clearly,
conditioned on no abort, the distribution of output is
$\p$. Furthermore, the algorithm declares $\bot$ if all the groups
declare $\bot$, which happens with probability at most
$(3/4)^{10\clg{\log\frac{1}{\alpha}}}<\alpha$.  
\end{proofof}

%% file: sec-simulation-inference.tex
\section{Simulate-and-Infer}\label{sec:simulate:infer}
We now show how to use distributed simulation results to design
private-coin distributed inference protocols. The approach is natural:
Simulate enough independent samples at the referee $\referee$ to solve
the centralized problem. We first describe 
the implications of the results from~\cref{sec:simulation}
for {any} distributed inference task, and then instantiate them
to our two specific applications: distribution learning and identity
testing.    

\subsection{Private-coin $\numbits$-bit distributed inference via distributed simulation}\label{sec:sampling:general:application}

Using the distributed simulation protocols of the previous section, we
can simulate one sample from $\p$ at the referee using about
$(\ab/2^\numbits)$ players. Then, to solve an inference task in the
distributed setting, the referee can simulate the number of
samples needed to solve the task in the centralized setting. The
resulting protocol will require a number of players roughly equal to
the sample complexity of the inference problem when the samples are
centralized times $\big(\ab/2^\numbits)$, the number of players
required to simulate each independent sample at the referee.  
We refer to protocols that first simulate samples from the underlying
distribution and then use a centralized inference algorithm at the
referee as \emph{simulate-and-infer} protocols. For concreteness,
we provide a formal description in~\cref{alg:SAI}. 
\begin{algorithm}[ht]
\begin{algorithmic}[1]
\Require Parameters $C$, $N$, $\ns=4CN\clg{\frac{\ab}{2^{\numbits}-1}}$
players observing one sample each from an unknown $\p$, and a
(centralized) estimator $e$ for $\task$ requiring $N$ samples
\State Partition the players into blocks of size $4\clg{\frac{\ab}{2^{\numbits}-1}}$.
\State Execute instances of the distributed simulation protocol given in~\cref{alg:simulation-ell} on each block.
 \If {at least $N$ instances return (independent) samples $\hat{X} \neq \bot$,}
{take a subset $(\hat{X}_1, \dots, \hat{X}_{N})$ of these samples and output $\hat e = e(\hat{X}_1, \dots, \hat{X}_{N})$;}
\Else{ output an arbitrary element $\hat{e}\in\cE$.}
\EndIf
\end{algorithmic}
\caption{The simulate-and-infer protocol for $\task=(\class, \cX, \cE,\loss)$}
\label{alg:SAI}
\end{algorithm}
For $\vec{\gamma}\in \R_+^q$, let
$\psi_{\task}(\vec{\gamma})$ denote the sample complexity for
$\vec{\gamma}$-inference protocol to solve $\task$ in the centralized
setting. That is, $\psi_{\task}(\vec{\gamma})$ denotes the \postrevision{smallest} $\ns$ for which there exists an estimator $e$ such that for every $\p\in\class$ and $\ns$ independent samples from $\p$, we have
\[
L_i(\p, e) \leq \gamma_i, \quad\, \forall 1\leq i \leq q\,,
\]
where $L\in\R_+^q$ is defined in~\eqref{eq:loss}. 
The next result evaluates the performance of~\cref{alg:SAI}. 
\begin{theorem}
  \label{theo:exactsampling:implies:ub}
Let $\task=(\class, \cX, \cE,\loss)$ be an inference problem with
\emph{bounded} loss $\loss\colon\class\times\cE\to\R^q$; \ie,
$\norminf{\loss}\leq 1$. For $0<\delta$, $1\le \numbits\le \clg{\log \ab}$, and $\vec{\gamma}\in \R_+^q$, upon setting $N=\psi_{\task}(\vec{\gamma})$ and $C=2+(1/\psi_{\task}(\vec{\gamma}))\log(1/\delta)$, the simulate-and-infer protocol given in~\cref{alg:SAI} requires $\bigO{ (\psi_{\task}(\vec{\gamma})\vee \log\frac{1}{\delta})\cdot \frac{\ab}{2^{\numbits}} }$
players and constitutes an $\ell$-bit deterministic  $(\vec{\gamma}+\delta \mathbf{1}_q)$-inference protocol for $\task$.
\end{theorem}
\begin{proof}
   We denote the resulting distributed inference protocol by $(\pi,e')$, and proceed to show it is a $(\vec{\gamma}+\delta \mathbf{1}_q)$-inference protocol for $\task$. From~\cref{theo:generate:sample:lbits:1sample}, each block produces independently a sample with probability at least $1/4$ (and $\bot$ otherwise). Thus, by Hoeffding's inequality, the number of samples simulated is larger than $N=\psi_{\task}(\vec{\gamma})$ with probability at least $1-\delta$ as long as $(5C-1)^2/(10C)\geq {1}/{\psi_{\task}(\vec{\gamma})}\log({1}/{\delta})$, which is satisfied for $C\geq 2+({1}/{\psi_{\task}(\vec{\gamma})})\log({1}/{\delta})$. Denoting by $\cE$ the event that the referee can simulate at least $\psi_{\task}(\vec{\gamma})$ samples, the expected loss satisfies  
   \begin{align*}
        L_i(\p, e') %
        &\leq (1-\delta)\expectCond{\loss_i(\p,\hat{e})}{\cE} + \delta\expectCond{\loss_i(\p,\hat{e})}{\bar{\cE}}\\
        &\leq \expectCond{\loss_i(\p,\hat{e})}{\cE} + \delta\norminf{\loss_i}
\\        
&\leq L_i(\p, e) + \delta 
\\       
 &\leq \gamma_i + \delta,
   \end{align*}
   for every $1\leq i \leq q$, concluding the proof.  
\end{proof}
The theorem above is quite general and only requires that the loss function be bounded.\footnote{In particular, it is immediate to extend it to the more general bounded case $\norminf{\loss} < \infty$, instead of $\norminf{\loss} \leq 1$.} Further, it is worth noting that the dependence on $\delta$ is very mild and can even be ignored, for instance, in settings when $\vec{\gamma} = \gamma\textbf{1}_q$ with $\gamma \asymp \delta$ and $\psi_{\task}(\vec{\gamma}) \gtrsim \log(1/\delta)$ (as the next two examples will illustrate).

\subsection{Application: private-coin protocols from distributed simulation}\label{sec:sampling:applications}

As corollaries of~\cref{theo:exactsampling:implies:ub}, we obtain distributed
inference protocols for distribution learning and identity testing.

\noindent Using the well-known result\footnote{This can be shown, for instance, by considering 
the empirical distribution $\hat \p$ and using McDiarmid's inequality
to bound the probability of error event
$\{\totalvardist{\p}{\hat \p}>\dst\}$.} that
$\bigTheta{(\ab+\log(1/\delta))/\dst^2}$ samples are sufficient to
learn a distribution over $[\ab]$ to within a total variation distance
$\dst$ with probability $1-\delta$, we obtain the following.
\begin{corollary}\label{coro:exactsampling:implies:learning}
  For $\numbits\in\{1,\ldots, \clg{\log \ab}\}$, simulate-and-infer constitutes an
  $\numbits$-bit deterministic $(\ab, \dst, \delta)$-learning protocol
  with $\bigO{\frac{\ab}{2^\numbits \dst^2}(\ab+\log(1/\delta))}$
  players. In particular, for any constant $\delta\in(0,1]$,
  $O(\ab^{2}/2^\numbits\dst^2)$ players suffice. 
\end{corollary}

\noindent For identity testing, it is known that the sample complexity is $O((\sqrt{\ab\log(1/\delta)}+\log(1/\delta))/\dst^2)$ samples ($cf.$~\cite{HuangM13, DGPP:17}). Thus, we get the following corollary to~\cref{theo:exactsampling:implies:ub}. 
\begin{corollary}\label{coro:exactsampling:implies:testing:identity}
 For $\numbits\in\{1,\ldots, \clg{\log \ab}\}$, simulate-and-infer constitutes an
 $\numbits$-bit deterministic $(\ab, \dst, \delta)$-identity testing
 protocol with $\bigO{\frac{\ab}{2^\numbits \dst^2}(\sqrt{\ab\log(1/\delta)}+\log(1/\delta))}$ players. In particular, for any constant $\delta\in(0,1]$, $O(\ab^{3/2}/2^\numbits\dst^2)$ players suffice.
\end{corollary}
\begin{remark}\label{r:sample_complexity_private}
We highlight that for constant $\delta$, the two corollaries above are
known to be optimal among all private-coin protocols.  Indeed, up to
constant factors they achieve the sample complexity lower bounds
established in~\cite{ACT:18} for private-coin learning and uniformity
testing protocols, respectively. In particular, we remark that~\cref{coro:exactsampling:implies:testing:identity}
shows that simulate-and-infer attains the sample complexity
$\bigTheta{\ab^{3/2}/(2^\numbits \dst^2)}$ of identity testing using
  private-coin protocols. \postrevision{We leave establishing the optimality of our results with respect to the parameter $\delta$ as an interesting open question.}
\end{remark}
\subsection{Optimality of our distributed simulation protocol}\label{sec:sampling:optimality:simulation}
Interestingly, a byproduct of our performance bound for simulate-and-infer protocols
(more precisely, that of~\cref{coro:exactsampling:implies:learning})
is that the $\alpha$-simulation protocol
from~\cref{theo:generate:sample:lbits:1sample} has optimal number of
players, up to constants.
\begin{corollary}\label{coro:optimality:lasvegas}
  For $\numbits \in\{1,\ldots, \clg{\log \ab}\}$ and $\alpha \in
  (0,1)$, any $\numbits$-bit public-coin (possibly interactive)
  $\alpha$-simulation protocol for $\ab$-ary distributions must have
  $\ns=\Omega(\ab/2^\numbits)$ players.
\end{corollary}
\begin{proof}
  Let $\pi$ be any $\numbits$-bit $\alpha$-simulation protocol with
  $\ns$ players. Proceeding analogously to proofs of~\cref{theo:exactsampling:implies:ub} and~\cref{coro:exactsampling:implies:learning}, we get
  that $\pi$ can be used to get an $\numbits$-bit $(\ab, \dst, 1/3)$-learning
  protocol for $\ns' = \bigO{\ns\cdot {\ab}/{\dst^2}}$
  players. (Moreover, the resulting
  protocol is adaptive, private- or public-coin,
  respectively, whenever $\pi$ is.) However, as shown
  in~\cite{HOW:18} (see, also,~\cite{ACT:18}), any
  $\numbits$-bit public-coin (possibly interactive) $(\ab, \dst,
  1/3)$-learning protocol must have   $\bigOmega{\ab^2/(2^\numbits\dst^2)}$ players. It follows that $\ns$  must satisfy $\ns \gtrsim \ab/2^\numbits$, as claimed.
\end{proof}

%% file: sec-identity.tex
\section{Public-coin identity testing}\label{sec:identity}

In this section, we propose public-coin protocols for $(\ab,
\dst)$-identity testing and establish the following upper bound on the
number of players required.
\begin{theorem}\label{theo:identity:shared:randomness:ub}
For $1\leq \numbits \leq \clg{\log \ab}$, there exists an
$\numbits$-bit public-coin $(\ab, \dst)$-identity testing protocol for
$\ns=\bigO{\frac{\ab}{2^{\numbits/2}\dst^2}}$ players.
\end{theorem}

In view of~\cref{r:sample_complexity_private} and the previous result,
public-coin protocols require a factor $\sqrt{\ab/2^\numbits}$ fewer
samples than private-coin protocols for identity testing. To the best
of our knowledge, this is one of the first instances of a natural
distributed inference problem where the availability of public coins
changes the sample complexity.  In fact, it follows from~\cite{ACT:18}
that the sample requirement of
$\bigO{\frac{\ab}{2^{\numbits/2}\dst^2}}$
in~\cref{theo:identity:shared:randomness:ub} is optimal among all
public-coin protocols. Thus, our work provides sample optimal private-
and public-coin protocols for identity testing (the optimal bounds for
sample complexity are given in Table~\ref{table:results}).

We now present our public-coin protocol for distributed identity
testing that attains the bounds
of~\cref{theo:identity:shared:randomness:ub}.  The basic steps of our
scheme are the following:
\begin{enumerate}
\item We use the public coins for the players to agree on a {random partition} of the domain $[\ab]$ into $L\eqdef 2^\numbits$ parts $S_1, \ldots, S_L$ where $|S_j|=\ab/L$ for each $j$. 
\item Player $i$ then sends the message $Y_i$ to be the index $j\in[L]$ such that $X_i\in S_j$ using $\numbits$ bits. 
\end{enumerate}
We now elaborate on the two steps and their implications
below. Consider the set of all partitions of $[\ab]$ into $L$
parts of equal cardinalities;  we call such partitions
  balanced partitions. Each such partition $(S_1, \dots, S_L)$
corresponds to a mapping from $[\ab]$ to $[L]$, where
the pre-image of $j\in[L]$ corresponds to the set $S_j$, and exactly $\ab/L$
elements map to each $j$. Note that the number of such partitions is
given by $\binom{\ab}{\ab/L \ldots \ab/L}$. The players use public 
randomness to agree on one of these partitions uniformly at
random. For a distribution $\p\in\distribs{[\ab]}$ and a uniformly
chosen balanced partition $S_1, \ldots, S_L$, consider the
distribution induced over $[L]$ as follows:  
\begin{equation} 
  Z_r(\p) \eqdef \p(S_r)\,, \quad r\in [L],
  \label{e:partition_distribution}
\end{equation} 
where $\p(S_r)$ is the probability assigned to $S_r$ by $\p$. 

For two distributions $\p$ and $\q$ over $[\ab]$ we will show that
with a constant probability under the randomized partitions, the
distance between the $\p$ and $\q$ are {preserved} (up to a
constant factor) by the induced distributions $\overline{\p}=(Z_1(\p),\dots,
Z_L(\p))$ and $\overline{\q}=(Z_1(\q),\dots, Z_L(\q))$. If $\p=\q$,
then clearly $\overline{\p}=\overline{\q}$. We next prove that if $\p$
and $\q$ are far (in total variation distance), then the induced
distributions, too, are far (in $\lp[2]$ distance). 
\begin{theorem}\label{theo:identity:z:concentration:anticoncentration:general}
  Fix any $\ab$-ary distributions $\p,\q$. For the (random)
  distributions $\overline{\p}=(Z_1(\p),\dots, Z_L(\p))$,
  $\overline{\q}=(Z_1(\q),\dots, Z_L(\q))$ over $[L]$ defined in~\cref{e:partition_distribution} above, the following holds: (i)~if $\p=\q$, then 
  $\overline{\p}=\overline{\q}$ with probability one;
  and (ii)~if $\totalvardist{\p}{\q} > \dst$, then
  \[
    \probaOf{ \normtwo{\overline{\p}-\overline{\q}}^2 >
      \frac{\dst^2}{2\ab} } \geq c\,,
  \] for some absolute constant $c>0$. 
\end{theorem}
The proof of this result involves proving the anticoncentration of
$\sum_{r\in[L]} \left(\sum_{j\in[\ab]}(\p_j-\q_j)\indicSet{\{j\in S_r\}}
\right)^2$. Since the random variables $\indicSet{\{j\in S_r\}}$ are dependent, the analysis becomes
technical and requires analyzing the higher moments of the summation
above, before applying the Paley--Zygmund inequality. The complete proof is
deferred to~\cref{app:contracting:hashing}.

We now provide a sketch of the referee's algorithm for identity testing. By definition, the $\ns$ messages are independent and distributed according to $\overline{\p}$. When $\p=\q$, by the above $\overline{\p}=\overline{\q}$. When $\totalvardist{\p}{\q} > \dst$, however, with a constant probability (we will amplify the success probability later) we have that $\lp[2](\overline{\p}, \overline{\q})>\dst/\sqrt{2\ab}$. Therefore the problem at the referee is to test whether the samples are from a reference distribution $\overline{\q}$ over $[L]$ or at least $\dst/\sqrt{2\ab}$ in $\lp[2]$ distance. 

Consider first the special case of $\numbits=1$, and $\q=\uniformOn{\ab}$, namely uniformity testing with one bit communication. In this case, we have $L=2$ and $\overline{\q}=(1/2, 1/2)$ is a fair coin. It is well-known that the task of testing whether $\overline{\p}$ is a fair coin or if it has bias at least $\dst/\sqrt{\ab}$ requires $\Theta(1/(\dst/\sqrt{\ab})^2)=\Theta(\ab/\dst^2)$ samples. For comparison, note that in the private-coin case protocols required $\ab^{3/2}/\dst^2$ samples, and therefore this simple algorithm provides an improvement over them by a factor of $\sqrt{\ab}$. 
 
Turning to $\numbits > 1$, for the special case of testing uniformity (i.e., when $\q=\uniformOn{\ab}$), the referee observes realizations from a uniform random variable with values in $[L]$ when $\p=\uniformOn{\ab}$. However, when
$\totalvardist{\p}{\q} > \dst$, we only know that the observed $L$-ary
random variable has distribution that is $(\dst/\sqrt{\ab})$-far from the uniform distribution in 
$\lp[2]$ distance (with constant probability), and not $\dtv$ as above. We can however leverage~\cite[Proposition~3.1]{CDVV:14}
or~\cite[Theorem~2.10]{CDGR:17:journal}, which proposed a test for testing if an $L$-ary distribution is uniform or $(\gamma/\sqrt{L})$-far from uniform in $\lp[2]$ using $\bigO{\sqrt{L}/\gamma^2}$ samples. In our case, we want to test if the distribution is $\dst/\sqrt{\ab} = \dst{\sqrt{L/k}}/\sqrt{L}$ far from uniform in $\lp[2]$ distance. Setting $\gamma \eqdef \dst{\sqrt{L/k}}$ this yields an algorithm that requires $\bigO{\sqrt{L}/\gamma^2}=\bigO{\ab/(2^{\numbits/2}\dst^2)}$ samples (for $L=2^\numbits$), which is the  number of players promised by~\cref{theo:identity:shared:randomness:ub}.

The arguments above are for the special case where the reference distribution $\q$ is uniform.  For a general reference distribution $\q$, our approach first involves reducing identity testing for
$\q$ to uniformity testing. Towards this, we rely on the following
result of Goldreich~\cite{Goldreich:16}, which we state here for
completeness. 
\begin{lemma}\label{l:goldreich}
For any $\q\in \distribs{\ab}$, there exists a randomized mapping $F_\q:\distribs{\ab}\to
\distribs{5\ab}$ satisfying the following properties:~~(i) $F_\q(\q)=\uniformOn{5\ab}$;
~~(ii) for every $\p\in \distribs{\ab}$ such that
$\totalvardist{\p}{\q}\geq \dst$, it holds that  
$\totalvardist{F_\q(\p)}{\uniformOn{5\ab}}\geq 16\dst/25$; and~~(iii)
there is an efficient algorithm for generating a sample from
$F_\q(\p)$ given one sample from $\p$.
\end{lemma}
\begin{remark} The mapping $F_\q$ and the algorithm mentioned in property~(iii) above require the knowledge of $\q$. 
\end{remark}

With this result at our disposal, each player can simply simulate
samples from $F_\q(\p)$ when they observe samples from
$\p$. Thereafter we can simply apply the distributed uniformity test
we outlined earlier, however for a slightly inflated domain of
cardinality $5\ab$. 

Recall that in~\cref{theo:identity:z:concentration:anticoncentration:general}, when $\p=\q$  the distribution of messages is equal to $\overline{\q}$ with probability one, but when the distributions are far (i.e., $\lp[2](\overline{\p}, \overline{\q})>\dst/\sqrt{2\ab}$) with only a constant probability $c$. We will now  ``amplify'' these constant probabilities to our desired probability of $11/12$. In fact, the amplification technique we present, considered folklore in the computational learning community, allows us to amplify easily the probabilities to any arbitrary $\delta$. We summarize this simple amplification in the next result.
\begin{lemma}\label{l:amplify}
For $\theta_1>1-\theta_2$, consider $N$ independent samples generated from $\bernoulli{p}$ with
either $p\geq \theta_1$ or $p\leq 1-\theta_2$. Then, for
$N=\bigO{1/(\theta_1+\theta_2-1)^2\log 1/\delta}$, we can find a test
that accepts $p\geq \theta_1$ with probability greater than $1-\delta$
in the first case and rejects it with probability greater than
$1-\delta$ in the second case.
\end{lemma}
The test is simply the empirical average with an appropriate threshold
and the proof follows from a standard Chernoff bound. We omit the
details. 

As a corollary of~\cref{l:amplify}
and~\cref{theo:identity:shared:randomness:ub}, we obtain the following result.
\begin{corollary}\label{c:identity:shared:randomness:ub:delta}
For $1\leq \numbits \leq \clg{\log \ab}$, there exists an $\numbits$-bit
public-coin $(\ab, \dst,\delta)$-identity testing protocol for
$\ns=\bigO{\frac{\ab}{2^{\numbits/2}\dst^2}\log\frac{1}{\delta}}$
players.
\end{corollary}
\begin{proof}
Recall that by our definition of $(\ab, \dst)$-identity testing
and~\cref{theo:identity:shared:randomness:ub}, we are given a test
with probability of correctness greater than $11/12$. Thus, when
$\p=\q$, the referee's output bit takes value $1$ with probability
exceeding $11/12$ and when $\totalvardist{\p}{\q}\geq \dst$, the
output bit takes value $0$ with probability exceeding $11/12$. Therefore,
the claimed test in the statement of the corollary is obtained by
applying the test of~\cref{theo:identity:shared:randomness:ub} to
$\bigO{\log 1/\delta}$ blocks of 
$\bigO{\ab/2^{\numbits/2}\dst^2}$ players and applying the
 test in~\cref{l:amplify} to the binary outputs of these
 tests. 
\end{proof}

We summarize our overall distributed identity test
in~\cref{alg:public-coin-IT} below. 
\begin{algorithm}[ht]
\begin{algorithmic}[1]
\Require Parameters $\gamma\in (0,1)$, $N$, $\ns$
players observing one sample each from an unknown $\p$

\State\label{step:goldreich} Players use the algorithm in~\cref{l:goldreich} to convert
their samples from $\p$ to independent samples $\tilde{X}_1, \dots,
\tilde{X}_\ns$ from $F_\q(\p)\in \distribs{5\ab}$.\Comment{This step 
  uses only private randomness.}

\State Partition the players into $N$ blocks of size $m\eqdef\ns/N$.
\State Players in each block use independent public coins to sample a
random partition $(S_1,\dots, S_L)$ with equal-sized parts. We represent
this partition by $(Y_1, \dots, Y_{5\ab})$ with $Y_r\in [L]$ as mentioned
above. 

\State Upon observing the sample $\tilde{X}_j=i$
in~\cref{step:goldreich}, player $j$ sends $Y_i$ (corresponding to
  its respective block) represented by  $\numbits$ bits. 

\State\label{step:center_l2}For each block, the referee obtains $\ns/N$ independent samples
from $(Z_1(\p), \dots, Z_L(\p))$ and tests if the underlying
distribution is $\uniformOn{L}$ or $(\gamma/\sqrt{L})$-far from uniform
in $\lp[2]$, with failure probability $\delta' \eqdef c/2(1-c)$.
\Comment This uses the aforementioned test
from~\cite{CDVV:14,CDGR:17:journal}; $c>0$ is as in~\cref{theo:identity:z:concentration:anticoncentration:general}.
\State The referee applies the test from~\cref{l:amplify} to the $N$ outputs of the independent tests
(one for each block) and declares the output.
\end{algorithmic}
\caption{An $\numbits$-bit public-coin protocol for distributed identity testing for
  reference distribution $\q$.}
\label{alg:public-coin-IT}
\end{algorithm}

We now show that with appropriate choice of
parameters,~\cref{alg:public-coin-IT} attains the performance promised
in~\cref{theo:identity:shared:randomness:ub}. 

\noindent\textit{Proof of~\cref{theo:identity:shared:randomness:ub}}. 
Our proof rests on two
technical results pointed 
above:~\cref{theo:identity:z:concentration:anticoncentration:general}
and~\cref{l:goldreich}. Consider the distributed identity test given
in~\cref{alg:public-coin-IT}. First, by~\cref{l:goldreich}, for any reference
distribution $\q$ the samples obtained by the players
in~\cref{step:goldreich} are independent samples from
$\uniformOn{5\ab}$ when $\p=\q$ and from a distribution that is
$(16\dst/25)$-far from $\uniformOn{5\ab}$ in total variation distance
when $\totalvardist{\p}{\q}>\dst$. 

The samples $(\tilde{X}_1,\dots,
\tilde{X}_\ns)$ are then ``quantized'' to $\numbits$ bits in each
block. For each block of $m=\ab/N$ players, we can consider the
samples seen by the referee as $m$ independent samples from an unknown
distribution on $[L]$. By the previous observation and~\cref{theo:identity:z:concentration:anticoncentration:general}, the common
distribution of independent samples at the referee in each block is
either $\uniformOn{L}$ with probability $1$ when $\p=\q$, or $(\dst/{10\ab})$-far\footnote{The
extra factor of $5$ is from~\cref{l:goldreich}.} from
$\uniformOn{L}$ in $\lp[2]$ distance with probability greater than $c$.

We set  $\gamma\eqdef \dst\sqrt{L}/{\sqrt{10\ab}}$ and apply the test
from~\cite{CDVV:14} or~\cite{CDGR:17:journal}. The test will succeed
if the event in~\cref{theo:identity:z:concentration:anticoncentration:general} occurs
and the centralized uniformity test succeeds. By~\cite[Proposition~3.1]{CDVV:14}
or~\cite[Theorem~2.10]{CDGR:17:journal}, this happens with probability
greater than $(1-\delta^\prime)c$ if  the number of samples $m$ in each
block exceeds
\begin{equation}\label{eq:l2:testing}
    \frac{\sqrt{L}}{\gamma^2} = \frac{10\ab\sqrt{L}}{L\dst^2}
=    \frac{10\ab}{\sqrt{L}\dst^2}.
\end{equation} 
We set the number of players in each block as $m\eqdef\clg{ 10\ab/(2^{\numbits/2}\dst^2) }$. Note that the parameter
$\delta^\prime$ here is the chosen probability of failure of the centralized
test. For our purpose, we shall see that it suffices to set it to $\delta^\prime \eqdef c/2(1+c)$.

Each block now provides a uniformity test which succeeds with
probability exceeding $1-\delta^\prime=(1+c/2)/(1+c)$. Finally, we amplify the probability of
success by choosing the number of blocks $N$ to be appropriately
large. We do this using the general amplification given in~\cref{l:amplify}.
Specifically, when $\p=\q$, the test for each of the block outputs $1$
with probability greater than $1-\delta^\prime=(1+c/2)/(1+c)$. On the other
hand, when $\p$ is $\dst$-far from $\q$, the test for each 
block outputs $0$ with probability greater than $(1-\delta^\prime)c =
(c+c^2/2)/(1+c)$. Therefore, the claim follows upon applying~\cref{l:amplify} with
$\theta_1\eqdef (1+c/2)/(1+c)$ and $\theta_2\eqdef (c+c^2/2)(1+c)$, which satisfy
$\theta_1>1-\theta_2$. \qed

Note that the protocol
  in~\cref{alg:public-coin-IT} is
  remarkably simple, and, moreover, is ``smooth,'' in the sense that
  no player's output depends too much on any particular symbol from
  $[\ab]$. (Indeed, each player's output is the indicator of a set of
  $\ab/2^\numbits$ elements, which for constant values of $\numbits$
  is $\Omega(\ab)$.) This ``smoothness'' can be a desirable feature
  when applying such protocols on a distribution whose domain
  originates from a quantization of a larger or even continuous
  domain, where the output of the test should not be too sensitive to
  the particular choice of quantization. Moreover, it is worth noting that the knowledge of the shared randomness
  by the referee is not used in~\cref{alg:public-coin-IT}.

\begin{remark}[Amount of shared randomness]
  \label{rk:randomness:used}
It is easy to see that~\cref{alg:public-coin-IT} uses no more than
$\bigO{\numbits \ab}$ bits of shared randomness. Indeed, 
$N=\Theta(1)$ independent partitions of $[\ab]$ into
$L\eqdef 2^\numbits$ equal-sized parts are chosen and each such partition can be
specified using $O(\log(L^\ab)) = O(\ab\cdot \numbits)$ bits.
  As mentioned in the preceding discussion, the proof of~\cref{theo:identity:shared:randomness:ub} hinges
  on~\cref{theo:identity:z:concentration:anticoncentration:general},
  whose proof relies in turn on an anticoncentration argument only involving
  moments of order four or less of suitable random variables. 
As such,
  one could hope that using $4$-wise independence (or a related
  notion) to sample the random equipartition of $[\ab]$ may lead to
  drastic savings in the number of shared random bits required to
  implement the protocol. 

This is indeed the case, with a caveat:
  namely, a straightforward way to
  implement~\cref{theo:identity:z:concentration:anticoncentration:general}
  would be to require a $4$-wise independent family of permutations of
  $[\ab]$ (see, \eg,~\cite{KNR:09,AL:13}).\footnote{Specifically,
    given such a family $\mathcal{F}$, one can obtain an equipartition
    of $[\ab]$ in $L$ pieces meeting our requirements by first fixing
    any equipartition $\Pi$ of $[\ab]$ in $L$ pieces, then drawing a
    permutation $\sigma\in\mathcal{F}$ uniformly at random, with $\log
    |\mathcal{F}|$ independent uniformly random bits, and applying
    $\sigma$ to $\Pi$.}{} Unfortunately, no non-trivial $t$-wise
  independent family of permutations is known to exist for $t>3$
  (although their existence is not ruled out).  A way to circumvent
  this issue and obtain a time- and randomness-efficient protocol
  using $O(\log \ab)$ shared random bits, is instead to observe
  that~\cref{theo:identity:z:concentration:anticoncentration:general}
  still holds for a uniformly random {partition} (instead of
  equipartition) of $[\ab]$ in $L$ pieces. This is because its
  proof invokes~\cref{theo:contraction:hashing:general}, which only
  requires suitable $4$-symmetric random variables. An efficient
  implementation then can rely on a family of $\ab$ $4$-wise
  independent random bits, for which explicit constructions with a
  seed length $O(\log \ab)$ are known. However, this approach hits another stumbling block, as when $\p=\q$ the resulting distribution $(Z_1(\q),\dots,Z_L(\q))$ on $[L]$ need not be uniform (as the partition is no longer in equal-sized parts), and thus the sample complexity from~\eqref{eq:l2:testing} (which holds for {uniformity} testing in $\lp[2]$ distance) does not follow. We explain in~\cref{app:randomness:efficiency} how to circumvent this difficulty and obtain a variant of~\cref{theo:identity:shared:randomness:ub} using only $O(\log \ab)$ shared random bits.
\end{remark}

\begin{remark}[Instance-optimal testing]
It may be of independent interest to consider instance-optimal
identity testing in the sense of Valiant and
Valiant~\cite{VV:14}, namely to examine how the number of players needed
depend on $\q$ instead of the worst-case parameter $\ab$. Towards that, we describe an extension of Goldreich's
reduction in~\cref{app:identity:from:uniformity} which makes it
amenable to the instance-optimal setting, and we believe will find further applications. 
\end{remark}

%% file: app-simulation-impossibility.tex
\subsection{Impossibility of perfect simulation in the interior of the probability simplex}\label{app:simulation:impossibility}
 
In this appendix, we establish~\cref{theo:sampling:impossibility:non:adaptive:k3}, restated below:
\begin{theorem}
For any $\ns\geq 1$, there does not exist any $\numbits$-bit perfect
simulation of ternary distributions ($\ab=3$) unless $\numbits\geq 2$,
even under when the input distribution is known to comes from an open
set in the interior of the probability simplex.
\end{theorem} 
Before we prove the theorem, we show that there is no loss of
generality in restricting to deterministic protocols, namely
protocols where each player uses a deterministic function of their
observation to communicate. The high-level argument is relatively simple: By
replacing player $j$ by two players $j_1,j_2$, each with a suitable
deterministic strategy, the two $1$-bit messages received by the
referee will allow it to simulate player $j$'s original randomized
mapping. A similar derandomization was implicit in~\cref{alg:simulation-ell-1-enhanced}. 

\begin{lemma}\label{lemma:sampling:impossibility:non:adaptive:deterministic}
For $\cX=\{0, 1, 2\}$, suppose there exists a $1$-bit perfect
simulation $S'=(\pi', \delta')$ with $\ns$ players. Then, we can find a $1$-bit perfect deterministic
simulation $S=(\pi,\delta)$ with $2\ns$ players such that, for each
$j\in [2\ns]$, the communication $\pi_j$ sent by player $j$ is a deterministic function of the sample $x_j$ seen by player $j$, \ie,
\[
\pi_j(x, u) =\pi_j(x), \qquad x\in \cX\,.
\]
\end{lemma}
\begin{proof}
Consider the mapping $f\colon\{0,1,2\}\times \{0,1\}^\ast\to \{0,1\}$. We will
show that we can find mappings $g_1\colon \{0,1,2\}\to \{0,1\}$,
$g_2\colon \{0,1,2\}\to \{0,1\}$, and
$h\colon \{0,1\}\times \{0,1\} \times \{0,1\}^\ast\to \{0,1\}$ such that for every
$u$ 
\begin{equation}\label{e:deterministic}
\probaOf{f(X, u)=1} = \probaOf{h(g_{1}(X_1), g_{2}(X_2),u)=1},
\end{equation}
where random variables $X_1$, $X_2$ take values in $\{0,1,2\}$ and are independent and identically distributed, with same distribution as $X$. We can then use 
this construction to get our claimed simulation $S$
Using $2\ns$ players as follows: Replace the communication $\pi_j'(x,u)$ from  player
$j$ with communication  $\pi_{2j-1}(x_{2j-1})$ and
$\pi_{2j}(x_{2j})$, respectively, from two players
$2j-1$ and $2j$, where $\pi_{2j-1}$ and $\pi_{2j}$ correspond to
mappings $g_1$ and $g_2$ above for $f= \pi'_j$. The referee can then
emulate the original protocol using the corresponding mapping $h$ and
using $h(\pi_{2j-1}(x_{2j-1}), \pi_{2j}(x_{2j}), u)$ in place of communication
from player $j$ in the original protocol. Then, since the probability
distribution of the communication does not change, we retain the
performance of $S'$, but using only deterministic communication now.

Therefore, it suffices to establish \eqref{e:deterministic}. For convenience, denote $\alpha_u\eqdef \indic{f(0,u) =
1}$, $\beta_u\eqdef \indic{f(1,u) = 1}$, and
$\gamma_u\eqdef \indic{f(2,u) = 1}$. Consider the case when at most one of $\alpha_u, \beta_u, \gamma_u$ is $1$. In this case, we can assume without loss of generality that
$\alpha_u\leq \beta_u+\gamma_u$ and $(\beta_u+\gamma_u-\alpha_u)\in\{0,1\}$. Let $g_i(x)=\indic{x=i}$ for
$i\in\{1,2\}$. Consider the mapping $h$ given by
\[
h(0,0,u)=\alpha_u,\,\, h(1,0,u)=\beta_u, \,\, h(0,1,u)=\gamma_u,\,\, h(1,1,u)=(\beta_u+\gamma_u-\alpha_u)\,.
\]
Then, for every $u$,
\begin{align*}
&\probaOf{h(g_1(X_1), g_2(X_2),
u)=1}
\\
&\qquad= \alpha_u(1-\p_1)(1-\p_2)+ \beta_u(1-\p_1)\p_2
+\gamma_u\p_1(1-\p_2)+(\beta_u+\gamma_u-\alpha_u)\p_1\p_2
\\
&\qquad= \alpha_u(1-\p_1-\p_2)+\beta_u\p_2+\gamma_u\p_1 = \probaOf{f(X,u) = 1}\,,
\end{align*}
which completes the proof for this case. For the other case, we can simply consider $(1-\alpha_u), (1-\beta_u)$, and $(1-\gamma_u)$ and proceed as in the case above to conserve $\probaOf{h(g_1(X_1), g_2(X_2), u)=0}$.
\end{proof}
We now prove~\cref{theo:sampling:impossibility:non:adaptive:k3}, but
in view of our previous observation, we only need to consider deterministic communication. 
\begin{proofof}{\cref{theo:sampling:impossibility:non:adaptive:k3}}
Suppose by contradiction that there exists such a $1$-bit deterministic perfect
simulation protocol $S=(\pi,\delta)$ for $\ns$ players on
$\cX=\{0,1,2\}$ such that $\pi_j(x,u)=\pi_j(x)$ for all $x$. Assume that this protocol is correct for all distributions $\p$ in the neighborhood of some $\p^\ast$ in the interior of the simplex. Consider a 
partition the players into three sets $\cS_0$, $\cS_1$, and $\cS_2$, with
\[
    \cS_i \eqdef \setOfSuchThat{ j\in[\ns] }{ \pi_j(i) = 1 }, \qquad  i\in\{0,1,2\}\,.
\]
Note that for deterministic communication the message $M$ is
independent of public randomness $U$. Then, by the definition of perfect simulation, it must be the case that
\begin{align}\label{eq:lowerbound:sampling:output-probability}
\p_x &= \shortexpect_{U} \sum_{m\in\{0,1\}^{\ns}} \delta_x(m,U) \probaCond{ M = m }{ U } = \shortexpect_{U} \sum_{m} \delta_x(m,U) \probaOf{ M = m } \notag\\
     &= \sum_{m} \shortexpect_U[\delta_x(m,U)] \probaOf{ M = m }
\end{align}
for every $x\in\cX$, which with our notation of $\cS_0, \cS_1, \cS_2$
can be re-expressed as
\begin{align*}
\p_x &= \sum_{m\in\{0,1\}^{\ns}} \shortexpect_U[\delta_x(m,U)] \prod_{i=0}^2 \prod_{j\in\cS_i} (m_j \p_i + (1-m_j)(1-\p_i)) \\
    &= \sum_{m\in\{0,1\}^{\ns}} \shortexpect_U[\delta_x(m,U)] \prod_{i=0}^2 \prod_{j\in\cS_i} (1-m_j +(2m_j-1)\p_i)\,,
\end{align*}
for every $x\in\cX$. But since the right-side above is a polynomial in
$(\p_0,\p_1, \p_2)$, it can only be zero in an open set in the interior
if it is identically zero. In particular, the constant term must be zero:
\begin{align*}
    0
    = \sum_{m\in\{0,1\}^{\ns}} \shortexpect_U[\delta_x(m,U)] \prod_{i=0}^2 \prod_{j\in\cS_i} (1-m_j)
    = \sum_{m\in\{0,1\}^{\ns}} \shortexpect_U[\delta_x(m,U)] \prod_{j=1}^\ns
    (1-m_j)\,.
\end{align*}
Noting that every summand is non-negative, this implies that for all
$x\in\cX$ and $m\in\{0,1\}^\ns$,
\[
\shortexpect_U[\delta_x(m,U)] \prod_{j=1}^\ns (1-m_j) = 0.
\]
 In particular, for the all-zero message $\textbf{0}^\ns$, we get
$ \shortexpect_U[\delta_x(\textbf{0}^\ns,U)] = 0 $ for all $x\in\cX$,
so that again by non-negativity we must have
$\delta_x(\textbf{0}^\ns,u)=0$ for all $x\in\cX$ and randomness
$u$. But the message $\textbf{0}^\ns$ will happen with probability
\[
\probaOf{M=\textbf{0}^\ns} = \prod_{i=0}^2 \prod_{j\in\cS_i} (1-\p_i)
= (1-\p_0)^{\abs{\cS_0}}(1-\p_1)^{\abs{\cS_1}}(1-\p_2)^{\abs{\cS_2}} > 0,
\]
where the inequality holds since $\p$ lies in the interior of the
simplex. Therefore, for the output $\hat{X}$ of the referee we have
\begin{align*}
    \probaOf{ \hat{X} \neq \bot }
    &= \sum_m \sum_{x\in \cX} \shortexpect_U[\delta_x(m,U)]\cdot \probaOf{M=m}
    = \sum_{m\neq \textbf{0}^\ns} \probaOf{M=m} \sum_{x\in \cX} \shortexpect_U[\delta_x(m,U)] \\
    &\leq \sum_{m\neq \textbf{0}^\ns} \probaOf{M=\textbf{0}^\ns} =
    1- \probaOf{M=\textbf{0}^\ns} < 1\,,
\end{align*}
contradicting the fact that $\pi$ is a perfect simulation protocol.
\end{proofof}

\begin{remark}
It is unclear how to extend the proof
of~\cref{theo:sampling:impossibility:non:adaptive:k3} to arbitrary
$\ab, \numbits$. In particular, the proof of~\cref{lemma:sampling:impossibility:non:adaptive:deterministic}
does not extend to the general case. A plausible proof-strategy is a  
black-box application of the $\ab=3$, $\numbits=1$ result 
to obtain the general result  using a direct-sum-type argument.
\end{remark}

%% file: app-contracting-hashing.tex
\subsection{Proof of~\cref{theo:identity:z:concentration:anticoncentration:general}}\label{app:contracting:hashing}
In this appendix, we
prove~\cref{theo:identity:z:concentration:anticoncentration:general},
stating that taking a random balanced partition of the domain in
$L\geq 2$ parts preserves the $\lp[2]$ distance between distributions
with constant probability. Note that the special case of $L=2$
  was proven in the extended abstract~\cite{ACFT:18}, in a similar
  fashion. 

We begin by recalling the Paley--Zigmund inequality, a key tool we
shall rely upon.
\begin{theorem}[Paley--Zygmund]\label{theo:paley:zygmund}
    Suppose $U$ is a non-negative random variable with finite
    variance. Then, for every $\theta\in[0,1]$,
    \begin{equation*}
        \probaOf{ U > \theta\expect{U} } \geq
        (1-\theta)^2\frac{\expect{U}^2}{\expect{U^2}}\,.
    \end{equation*}
\end{theorem}
We will prove a more general version
of~\cref{theo:identity:z:concentration:anticoncentration:general},
showing that the $\lp[2]$ distance to any fixed distribution
$\q\in\distribs{[\ab]}$ is preserved with a constant
probability\footnote{For this application, one should read the theorem
  statement with $\delta \eqdef \p-\q$.} with only mild assumptions on
$Y_1,\dots,Y_\ab$; recall that we represent the partition $(S_1,\dots, S_L)$ using a $\ab$-length vector $(Y_1, \dots, Y_\ab)$ with each $Y_i\in [L]$ such that $Y_i=j\in[L]$ if $i\in S_j$. Namely, we only require that they be \emph{4-symmetric}:
\begin{definition}
  Fix any $t\in\N$. The random variables $Y_1,\dots,Y_\ab$ over $\Omega$ are said to be \emph{$t$-symmetric} if, for every $i_1,i_2,\dots,i_t\in[\ab]$, every $s\in\N$, and $f_1,\dots,f_s\colon \Omega^t \to \R$, the expectation $\bEE{\prod_{j=1}^s f_j(Y_{i_1},\dots,Y_{i_t})}$ may only depend on the multiset $\{i_1,i_2,\dots,i_t\}$ via its multiplicities. That is, for every permutation $\pi\colon[\ab]\to[\ab]$, 
\[
    \bEE{\prod_{j=1}^s f_j(Y_{i_1},\dots,Y_{i_t})} = \bEE{\prod_{j=1}^s f_j(Y_{\pi(i_1)},\dots,Y_{\pi(i_t)})}\,.
\]
\end{definition}
Before stating the general statement we shall establish, we observe that random variables $Y_1,\dots, Y_\ab$ as in~\cref{theo:identity:z:concentration:anticoncentration:general} are indeed $t$-symmetric for any $t\in[\ab]$. Another prominent example of $t$-symmetric random variables is that of independent, or indeed $t$-wise independent, identically distributed r.v.'s (and indeed, it is easy to see that $t$-symmetry for $t\geq 2$ require that the random variables be identically distributed). Moreover, for intuition, one can note that for $\Omega=\{0,1\}$, the definition amounts to asking that the expectation $\bEE{\prod_{s=1}^t Y_{i_s}}$ depends only on the multiplicities of the multiset $\{i_1,i_2,\dots,i_t\}$. 
\begin{theorem}[Probability Perturbation Hashing]
  \label{theo:contraction:hashing:general}
  Suppose $2\leq L< \ab$ is an integer dividing $\ab$, and fix any vector $\delta\in\R^\ab$ such that $\sum_{i\in[\ab]} \delta_i =0$. Let random variables $Y_1,\dots,Y_\ab$ be $4$-symmetric r.v.'s. Define $Z=(Z_1,\dots,Z_L)\in \R^L$ as
    \begin{equation*}
        Z_r \eqdef \sum_{i=1}^\ab \delta_i \indic{Y_i=r},\qquad r\in[L]\,.
    \end{equation*}
  Then, for every $\alpha\in(0,1/2)$,
  \begin{equation*}
      \probaOf{ \probaOf{Y_1\neq Y_2}- 4\sqrt{2\alpha} \leq \frac{\normtwo{Z}^2}{\normtwo{\delta}^2} \leq \min\!\Paren{\frac{4}{\sqrt{\alpha}}, \frac{\probaOf{Y_1\neq Y_2}}{\alpha}} } \geq \alpha\,.
  \end{equation*}
\end{theorem}%

\begin{proofof}{\cref{theo:contraction:hashing:general}}
The gist of the proof is to consider a suitable non-negative random variable (namely, $\normtwo{Z}^2$) and bound its
expectation and second moment in order to apply the Paley--Zygmund
inequality to argue about anticoncentration around the mean. The
difficulty, however, lies in the fact that bounding the moments of
$\normtwo{Z}$ involves handling the products of correlated $L$-valued
random variables $Y_i$'s, which is technical even for the case $L=2$
considered in~\cite{ACFT:18}. For ease of presentation, we have divided the argument into smaller results. 

In what follows, let random variables $Y_1,\dots, Y_\ab$ be as in the statement. Since they are $4$-symmetric, expectations of the form $\expect{f(Y_a,Y_b,Y_c,Y_d)g(Y_a,Y_b,Y_c,Y_d)}$ depend only on the number of times each distinct element appears in the multiset $\{a,b,c,d\}$. For ease of notation, we introduce the quantities below, for $r_1,r_2,r_3\in[L]$ (not necessarily distinct):\footnote{We assume throughout that $\ab\geq 4$. This is without loss of generality, as all results in this paper hold trivially for constant $\ab$.}
\begin{align*}
m_r &\eqdef \probaOf{Y_1 = r}\,,\\
m_{r_1,r_2} &\eqdef \probaOf{Y_1 = r_1,Y_2 = r_2}\,,\\
m_{r_1,r_2,r_3} &\eqdef \probaOf{Y_1 = r_1,Y_2 = r_2,Y_3 = r_3}\,,\\
m_{r_1,r_2,r_3,r_4} &\eqdef \probaOf{Y_1 = r_1,Y_2 = r_2,Y_3 = r_3,Y_4 = r_4}\,.
\end{align*}
With this notation at our disposal, we are ready to proceed with the proof. 

\begin{lemma}[Each part has the right expectation]\label{lemma:expectation:z:general:vanilla}
  For every $r\in[L]$, 
  \[
      \expect{Z_r} = 0\,.
  \]
\end{lemma}
\begin{proof}
By linearity of expectation, for every $r$,
$
    \expect{Z_r}  = \sum_{i=1}^\ab \delta_i  \expect{ \indic{Y_i=r} } = m_r \cdot \sum_{i=1}^\ab \delta_i = 0
.
$
\end{proof}

\begin{lemma}[The {$\lp[2]^2$} distance has the right expectation]\label{lemma:variance:z:general:vanilla}
For every $r\in[L]$,
  \[
      \var Z_r = \expect{Z_r^2} = (m_r - m_{r,r})\normtwo{\delta}^2\,.
  \]
  In particular, the expected squared $\lp[2]$ norm of $Z$ is
  \begin{equation*}
      \expect{\normtwo{Z}^2 } = \expect{\sum_{r=1}^L Z_r^2 } %
      = \Paren{1-\sum_{r=1}^Lm_{r,r}}\normtwo{\delta}^2
      = \probaOf{Y_1\neq Y_2}\cdot \normtwo{\delta}^2\,.
  \end{equation*}
\end{lemma}
\begin{proof}
For a fixed $r\in[L]$, using the definition of $Z$, the fact that $\sum_{i=1}^\ab  \indic{Y_i=r}  = \frac{\ab}{L}$, and~\cref{lemma:expectation:z:general:vanilla}, we get that
  \begin{align*}
      \var[ Z_r ] 
      &= \expect{ Z_r^2 } 
      = \expect{ \left(\sum_{i=1}^\ab \delta_i \indic{Y_i=r} \right)^2 }
      = \sum_{1\leq i,j\leq \ab} \delta_i\delta_j \expect{\indic{Y_i=r}\indic{Y_j=r}} \\
      &= \sum_{i=1}^\ab \delta_i^2 \expect{ \indic{Y_i=r} } + 2\sum_{1\leq i<j \leq \ab} \delta_i\delta_j \expect{ \indic{Y_i=r}\indic{Y_j=r} }\\
      &= m_r\sum_{i=1}^\ab \delta_i^2 + m_{r,r}\cdot 2\sum_{1\leq i<j \leq \ab} \delta_i\delta_j\\
      &= m_r\sum_{i=1}^\ab \delta_i^2 + m_{r,r}\Paren{\sum_{i=1}^\ab \delta_i}^2 - m_{r,r}\sum_{i=1}^\ab \delta_i^2\\
      &= (m_r - m_{r,r})\normtwo{\delta}^2\,.
  \end{align*}
  The conclusion follows noting that $\sum_{r=1}^L m_r = 1$, $\sum_{r=1}^L m_{r,r} = \probaOf{Y_1 = Y_2}$.
\end{proof}
For the lower tail bound, we will derive a bound for $\expect{Z^4}$ and invoke as discussed above the Paley--Zygmund inequality. Note that the lower bound trivially holds whenever $\alpha > \frac{1}{32}\probaOf{Y_1\neq Y_2}^2$; thus, we hereafter assume $0 \leq \alpha \leq \frac{1}{32}\probaOf{Y_1\neq Y_2}^2$. We have:
\begin{lemma}[The {$\lp[2]^2$} distance has the required second moment]\label{lemma:fourth:moment:z:general:vanilla}
  There exists an absolute constant $C>0$ such that
  \begin{equation*}
      \expect{\normtwo{Z}^4} \leq C\normtwo{\delta}^4\,.
  \end{equation*}
  Moreover, one can take $C=16$.
\end{lemma}
\begin{proofof}{\cref{lemma:fourth:moment:z:general:vanilla}}
Expanding the square, we have
\begin{equation}\label{eq:square:expanding}
    \expect{\normtwo{Z}^4} = \expect{\mleft(\sum_{r=1}^L Z_r^2\mright)^2}
    = \sum_{r=1}^L \expect{Z_r^4} + 2\sum_{r<r'} \expect{Z_r^2Z_{r'}^2}
\end{equation}
We will bound both terms separately. For the first term, we have the next bound, analogous to~\cite[Equation~(21)]{ACFT:18}.
\begin{claim}\label{claim:fourth:moment:z:general:first:term}
  For every
  $r\in[L]$,
  \[
      \expect{Z_r^4} \leq 12m_r\normtwo{\delta}^4\,,
  \]
  and therefore
  \[
      \sum_{r=1}^L \expect{Z_r^4} \leq 12 \normtwo{\delta}^4\,.
  \]
\end{claim}
\begin{proof}
We will mimic the proof
of~\cref{lemma:variance:z:general:vanilla}. We first rewrite
\begin{align*}
    \expect{Z_r^4}
    &= \expect{ \left(\sum_{i=1}^\ab \delta_i \indic{Y_i=r} \right)^4 }
    = \sum_{1\leq
    a,b,c,d \leq \ab} \delta_a\delta_b\delta_c\delta_d\expect{ \indic{Y_a=r}\indic{Y_b=r}\indic{Y_c=r}\indic{Y_d=r}
    }\,.
\end{align*}
Using symmetry once again, since every term
$\expect{ \indic{Y_a=r}\indic{Y_b=r}\indic{Y_c=r}\indic{Y_d=r} }$ depends only
on the number of distinct elements in the multiset $\{a,b,c,d\}$, it will be equal to one of $m_r, m_{r,r}, m_{r,r,r}$, or $m_{r,r,r,r}$, and it suffices to keep track of the contribution of each of these four types of terms. From this, letting
$\Sigma_s \eqdef \sum_{\abs{\{a,b,c,d\}}=s} \delta_a\delta_b\delta_c\delta_d
$ for $s\in[4]$, we get that
\begin{equation} \label{eq:fourthmoment:sigmas}
    \expect{Z_r^4} = m_r\Sigma_1
    + m_{r,r}\Sigma_2
    + m_{r,r,r}\Sigma_3
    + m_{r,r,r,r}\Sigma_4\,.
\end{equation}
We will rely on the following technical result.
\begin{fact}\label{fact:sigma:1234}
For $\Sigma_1,\Sigma_2,\Sigma_3$, and $\Sigma_4$ defined as above, we
have
\begin{align*}
\Sigma_1 &= \normfour{\delta}^4\\
\Sigma_2 &= 3\normtwo{\delta}^4 - 7\normfour{\delta}^4\\
\Sigma_3  &= 12\normfour{\delta}^4 - 6\normtwo{\delta}^4\\
\Sigma_4 &= -(\Sigma_1+\Sigma_2+\Sigma_3) = 3\normtwo{\delta}^4-6\normfour{\delta}^4\,.
\end{align*}
\end{fact}
\begin{proofof}{\cref{fact:sigma:1234}}
We start by showing the last equality: ``hiding zero,'' we get
\[
    0 = \left(\sum_{i=1}^\ab \delta_i\right)^4 = \sum_{1\leq a,b,c,d\leq \ab} \delta_a\delta_b\delta_c\delta_d = \Sigma_1+\Sigma_2+\Sigma_3+\Sigma_4\,.
\]
thus it is enough to establish the stated expressions for $\Sigma_1,\Sigma_2,\Sigma_3$. The first equality is a direct consequence of the definition $\Sigma_1 = \sum_{i=1}^\ab \delta_i^4 = \normfour{\delta}^4$; as for the second, we can derive it from
\begin{align*}
    \Sigma_2 
    &= \sum_{\substack{1\leq a,b,c,d\leq \ab\\ \abs{\{a,b,c,d\}}=2}} \delta_a\delta_b\delta_c\delta_d 
    = 6\sum_{i < j} \delta_i^2\delta_j^2 + 4\sum_{i<j} (\delta_i\delta_j^3+\delta_i^3\delta_j)
\\
    &= 3\left( \left(\sum_{i=1}^\ab \delta_i^2 \right)^2 - \sum_{i=1}^\ab \delta_i^4 \right) + 4\sum_{i<j} (\delta_i\delta_j^3+\delta_i^3\delta_j) \\
    &= 3\normtwo{\delta}^4 - 3\normfour{\delta}^4 + 4\sum_{i<j} (\delta_i\delta_j^3+\delta_i^3\delta_j)
    = 3\normtwo{\delta}^4 - 7\normfour{\delta}^4\,,
\end{align*}
where the last equality was obtained by ``hiding zero'' once more:
\[
    0 = \sum_{i=1}^\ab \delta_i \sum_{i=1}^\ab \delta_i^3 = \sum_{1\leq i,j\leq \ab} \delta_i\delta_j^3 = \sum_{i=1}^\ab \delta_i^4 + \sum_{i<j} (\delta_i\delta_j^3+\delta_i^3\delta_j)\,.
\]
Finally, to handle $\Sigma_3$, we expand
\begin{align*}
    \Sigma_3 
    &= \sum_{\substack{1\leq a,b,c,d\leq \ab\\ \abs{\{a,b,c,d\}}=3}} \delta_a\delta_b\delta_c\delta_d 
    = 12\sum_{a<b<c}( \delta_a^2\delta_b\delta_c+\delta_a\delta_b^2\delta_c+\delta_a\delta_b\delta_c^2)
\end{align*}
and, once more hiding zero, we leverage the fact that
\[
    0 = \left(\sum_{i=1}^\ab \delta_i\right)^2 \sum_{i=1}^\ab \delta_i^2 = \sum_{i=1}^\ab \delta_i^4 + 2\sum_{i<j} \delta_i^2\delta_j^2 + 2\sum_{i<j} (\delta_i\delta_j^3+\delta_i^3\delta_j)
    + 2\sum_{a<b<c}( \delta_a^2\delta_b\delta_c+\delta_a\delta_b^2\delta_c+\delta_a\delta_b\delta_c^2)
\]
\ie,
\[
    2\sum_{a<b<c}( \delta_a^2\delta_b\delta_c+\delta_a\delta_b^2\delta_c+\delta_a\delta_b\delta_c^2)
    = -\left( \normfour{\delta}^4 +\left(\normtwo{\delta}^4-\normfour{\delta}^4\right) -2\normfour{\delta}^4  \right)
    = 2\normfour{\delta}^4 - \normtwo{\delta}^4\,.
\]
This leads to $\Sigma_3  = 12\normfour{\delta}^4 - 6\normtwo{\delta}^4$.
\end{proofof}
Combing~\eqref{eq:fourthmoment:sigmas} with the above fact, we get
\begin{align*}
    \expect{Z_r^4} 
    &= 
      (m_r- 7m_{r,r}+ 12m_{r,r,r}+6m_{r,r,r,r})\normfour{\delta}^4
    + 3(m_{r,r}- 2m_{r,r,r}+ m_{r,r,r,r})\normtwo{\delta}^4\\
    &\leq 
      (m_r + 5m_{r,r,r}+6m_{r,r,r,r})\normfour{\delta}^4
      + 3(m_{r,r}- m_{r,r,r})\normtwo{\delta}^4\\
    &\leq 
      (m_r + 3m_{r,r} + 2m_{r,r,r} +6m_{r,r,r,r})\normtwo{\delta}^4\\
    &\leq 12m_r\normtwo{\delta}^4\,.
\end{align*}
leveraging the inequalities $\normtwo{\delta}\leq \normfour{\delta}$ and $m_{r,r,r,r} \leq m_{r,r,r} \leq m_{r,r} \leq m_r$. 
\end{proof}
However, we need additional work
to handle the second term comprising roughly $L^2$ summands. In
particular, to complete the proof we show that
each summand in the second term is less than a constant factor times
$m_{r,r'}\normtwo{\delta}^4$.
\begin{claim}\label{claim:fourth:moment:z:general:second:term}
  We have
  \[
      \sum_{r<r'} \expect{Z_r^2Z_{r'}^2} \leq 2\probaOf{Y_1\neq Y_2}\cdot \normtwo{\delta}^4\,.
  \]
\end{claim}
\begin{proof}
Fix any $r\neq r'$. As before, we expand
\begin{align*}
    \expect{Z_r^2Z_{r'}^2}
    &= \expect{ \left(\sum_{i=1}^\ab \delta_i \indic{Y_i=r} \right)^2\left(\sum_{i=1}^\ab \delta_i \indic{Y_i=r'} \right)^2 } \\
    &= \sum_{1\leq a,b,c,d \leq \ab} \delta_a\delta_b\delta_c\delta_d\expect{ \indic{Y_a=r}\indic{Y_b=r}\indic{Y_c=r'}\indic{Y_d=r'} }\,.
\end{align*}
We will use $4$-symmetry once again to handle the terms 
$\expect{ \indic{Y_a=r}\indic{Y_b=r}\indic{Y_c=r}\indic{Y_d=r} }$. The key observation here
is that if $\{a,b\}\cap\{c,d\} \neq \emptyset$, then
$\indic{Y_a=r}\indic{Y_b=r}\indic{Y_c=r'}\indic{Y_d=r'} = 0$. This
will be crucial as it implies that the expected value can only be
non-zero if $\abs{\{a,b,c,d\}}\geq 2$, yielding an $m_{r,r'}$ dependence
for the leading term in place of $m_r$. 
 \begin{align}
    \expect{Z_r^2Z_{r'}^2}
    &= \sum_{\abs{\{a,b,c,d\}}=2} \delta_a^2\delta_b^2\expect{ \indic{Y_a=r}\indic{Y_b=r'}} 
\notag
\\
    &\qquad+\sum_{\abs{\{a,b,c,d\}}=3} \delta_a^2\delta_b\delta_c\expect{ \indic{Y_a=r}\indic{Y_b=r'}\indic{Y_c=r'}} \notag\\
    &\qquad+\sum_{\abs{\{a,b,c,d\}}=3} \delta_a\delta_b\delta_c^2\expect{ \indic{Y_a=r}\indic{Y_b=r}\indic{Y_c=r'}} \notag\\
    &\qquad+\sum_{\abs{\{a,b,c,d\}}=4} \delta_a\delta_b\delta_c\delta_d\expect{ \indic{Y_a=r}\indic{Y_b=r}\indic{Y_c=r'}\indic{Y_d=r'}}\,. 
\label{eq:cross:terms}
\end{align}
The first term, which we will show dominates, can be expressed as
\[
\sum_{\abs{\{a,b,c,d\}}=2} \delta_a^2\delta_b^2\expect{ \indic{Y_a=r}\indic{Y_b=r'}}
= m_{r,r'} \normtwo{\delta}^4\,.
\]
For the second and the third terms, noting that 
\[
 \sum_{\abs{\{a,b,c,d\}}=3} \delta_a^2\delta_b\delta_c = \sum_{1\leq a,b,c\leq \ab} \delta_a^2\delta_b\delta_c - \sum_{a\neq b}\delta_a^2\delta_b^2 - 2\sum_{a\neq b}\delta_a^3\delta_b
\]
with $\sum_{1\leq a,b,c\leq \ab} \delta_a^2\delta_b\delta_c = \left(\sum_{a=1}^\ab \delta_a^2\right)\left(\sum_{a=1}^\ab \delta_a\right)^2 = 0$, $\sum_{a\neq b}\delta_a^2\delta_b^2\leq \sum_{1\leq a,b\leq \ab} \delta_a^2\delta_b^2 = \normtwo{\delta}^4$, and $\sum_{a\neq b}\delta_a^3\abs{\delta_b}\leq \sum_{1\leq a,b\leq \ab}\delta_a^3\abs{\delta_b} \leq \norminf{\delta}\norm{\delta}_3^3\leq \normtwo{\delta}^4$, we get
\[
      -m_{r,r',r'}\normtwo{\delta}^4 \leq  \sum_{\abs{\{a,b,c,d\}}=3} \delta_a^2\delta_b\delta_c \expect{ \indic{Y_a=r}\indic{Y_b=r'}\indic{Y_c=r'}}  \leq m_{r,r',r'}\normtwo{\delta}^4\,.
\]
Finally, similar manipulations yield
\[
    -m_{r,r,r',r'}\normtwo{\delta}^4
    \leq \sum_{\abs{\{a,b,c,d\}}=4} \delta_a\delta_b\delta_c\delta_d\expect{ \indic{Y_a=r}\indic{Y_b=r}\indic{Y_c=r'}\indic{Y_d=r'}}
    \leq m_{r,r,r',r'}\normtwo{\delta}^4\,.
\]
Gathering all this in~\eqref{eq:cross:terms}, we get that there exists some absolute constant $C'>0$ such that
\begin{align*}
    \sum_{r<r'} \expect{Z_r^2Z_{r'}^2}
    &\leq \normtwo{\delta}^4\cdot \sum_{r<r'} \Paren{ m_{r,r'}+m_{r,r,r'}+m_{r,r',r'}+m_{r,r,r',r'} }\\
    &\leq 2\normtwo{\delta}^4\cdot 2\sum_{r<r'} m_{r,r'}
    = 2\normtwo{\delta}^4\cdot \Paren{ \sum_{r,r'} m_{r,r'} - \sum_{r} m_{r,r} }\\
    &= 2\normtwo{\delta}^4\cdot \Paren{ 1 - \probaOf{Y_1=Y_2} }
    = 2\probaOf{Y_1\neq Y_2}\cdot \normtwo{\delta}^4\,,
\end{align*}
where we recalled the definition of $m_{r,r'} = \probaOf{Y_1=r,Y_2=r'}$ to re-express the sums.
\end{proof}
The lemma follows by combining~\cref{claim:fourth:moment:z:general:first:term,claim:fourth:moment:z:general:second:term}.
\end{proofof}
We are now ready to establish~\cref{theo:contraction:hashing:general}. By~\cref{lemma:variance:z:general:vanilla,lemma:variance:z:general:vanilla,lemma:fourth:moment:z:general:vanilla}, we have $\expect{\normtwo{Z}^2} = \probaOf{Y_1\neq Y_2}\normtwo{\delta}^2$ and $\expect{\normtwo{Z}^4} \leq 16\normtwo{\delta}^4$. Therefore, by the Payley--Zygmund inequality (\cref{theo:paley:zygmund}) applied to $\normtwo{Z}^2$, for every $\theta\in[0,1]$,
\begin{equation*}
  \probaOf{ \normtwo{Z}^2 > \theta\probaOf{Y_1\neq Y_2}\normtwo{\delta}^2 } 
  \geq (1-\theta)^2\frac{\expect{\normtwo{Z}^2}^2}{\expect{\normtwo{Z}^4}} 
  \geq (1-\theta)^2\frac{\probaOf{Y_1\neq Y_2}^2}{16}\,.
\end{equation*}
Choosing 
\[
\theta = 1-\frac{4\sqrt{2\alpha}}{\probaOf{Y_1\neq Y_2}}\,,
\]
so that the RHS is $2\alpha$, concludes the proof for the lower tail. 

For the upper tail, it follows from Chebyshev's inequality and~\cref{lemma:fourth:moment:z:general:vanilla} that, for any $C>0$, 
\begin{equation*}%
\probaOf{\normtwo{Z}^2 > C \probaOf{Y_1\neq Y_2}\cdot \normtwo{\delta}^2} \leq \frac{16}{C^2 \probaOf{Y_1\neq Y_2}^2}
\end{equation*}
which is equal to $\alpha$ for $C\eqdef \frac{4}{\sqrt{\alpha}\probaOf{Y_1\neq Y_2}}$. We also have 
$\probaOf{\normtwo{Z}^2 > \alpha^{-1} \probaOf{Y_1\neq Y_2}\cdot \normtwo{\delta}^2} \leq \alpha$ by Markov's inequality, and combining the two yields
\begin{equation}\label{eq:contraction:hashing:general:ub}
\probaOf{\normtwo{Z}^2 \leq \min\!\Paren{\frac{4}{\sqrt{\alpha}}, \frac{\probaOf{Y_1\neq Y_2}}{\alpha}}\cdot \normtwo{\delta}^2} \geq 1-\alpha.
\end{equation}
The overall theorem follows by a union bound over the upper and lower tail events.
\end{proofof}
We conclude this appendix by showing how~\cref{theo:identity:z:concentration:anticoncentration:general} readily follows from~\cref{theo:contraction:hashing:general}.
\begin{proofof}{\cref{theo:identity:z:concentration:anticoncentration:general}}
Since the first item is immediate, it suffices to prove the second, which we do now. Recall that the random variables $Y_1,\dots,Y_\ab$ from the statement of~\cref{theo:identity:z:concentration:anticoncentration:general} are such that each $Y_i$ is marginally uniform on $[L]$, and $\sum_{i=1}^\ab \indic{Y_i=r} = \frac{\ab}{L}$ for every $r\in[L]$. In particular, $Y_1,\dots,Y_\ab$ are $4$-symmetric random variables, as we see below:
\[
    \probaOf{Y_1\neq Y_2} = 1 - \sum_{r=1}^L \expect{\indic{Y_1=r}\indic{Y_2=r}}
    = 1 - \frac{1}{L^2}\cdot \frac{\ab-L}{\ab-1}
    \geq 1 - \frac{1}{L^2} \geq \frac{3}{4}\,.
\] 
Further, a simple computation yields
  \begin{align*}
      \expect{ \indic{Y_1=r}\indic{Y_2=r} }
      &= \expect{ \expectCond{ \indic{Y_1=r}\indic{Y_2=r} }{ \indic{Y_2=r} } } 
      = \frac{1}{L}\probaCond{ Y_1=r}{Y_2 = r} 
\\
      &= \frac{1}{L}\probaCond{ Y_1 = r}{\sum_{i=1}^{\ab-1} \indic{Y_i=r} = \frac{\ab}{L}-1}
      = \frac{1}{L^2}\cdot \frac{\ab-L}{\ab-1}\,,
  \end{align*}
where the final identity uses symmetry, along with the observation that 
\[
\sum_{i=1}^{\ab-1} \expectCond{ \indic{Y_i=r} }{\sum_{j=1}^{\ab-1} \indic{Y_j=r} = \frac{\ab}{L}-1} = \frac{\ab}{L}-1\,.
\]
Therefore, applying~\cref{theo:contraction:hashing:general} for $\alpha \eqdef \frac{1}{82} < \frac{1}{2}\Paren{\frac{8\probaOf{Y_1\neq Y_2}-1}{32}}^2$, with $\delta\eqdef \p-\q$, we obtain
\begin{equation*}
      \probaOf{ \normtwo{\overline{\p}-\overline{\q}}^2 \geq \frac{1}{2}\normtwo{\p-\q}^2 } \geq \alpha\,,
  \end{equation*}
which yields the desired statement, since by the Cauchy--Schwarz inequality we have $\normtwo{\p-\q}^2 > \frac{4\dst^2}{\ab}$ whenever $\totalvardist{\p}{\q} > \dst$.
\end{proofof}

%% file: app-randomnessefficient-identity.tex
\subsection{A randomness-efficient variant of~\cref{theo:identity:shared:randomness:ub}}\label{app:randomness:efficiency} 
In this appendix, we describe how the protocol
underlying~\cref{theo:identity:shared:randomness:ub},~\cref{alg:public-coin-IT},  
can be modified to reduce the number of shared bits 
from the $O(\ab\numbits)$ required by~\cref{alg:public-coin-IT} to only
$O(\log\ab)$. 
\begin{theorem}\label{theo:identity:shared:randomness:ub:efficiency}
For $1\leq \numbits \clg{\log \ab}$, there exists an $\numbits$-bit
public-coin $(\ab, \dst)$-identity testing protocol for
$\ns=\bigO{\frac{\ab}{2^{\numbits/2}\dst^2}}$ players, using $O(\log\ab)$ public coins.
\end{theorem}
\begin{algorithm}[ht]
\begin{algorithmic}[1]
\Require Parameters $\gamma\in (0,1)$, $N$, $\ns$
players observing one sample each from an unknown $\p$

\State Players use the algorithm in~\cref{l:goldreich} to convert
their samples from $\p$ to independent samples $\tilde{X}_1, \dots,
\tilde{X}_\ns$ from $F_\q(\p)\in \distribs{5\ab}$.\Comment{This step 
  uses only private randomness.}

\State Partition the players into $N$ blocks of size $m\eqdef\ns/N$.
\State\label{step:partition:new} Players in each block use $4(\clg{\log(5\ab)}+\numbits)$ independent public coins to generate (using~\cref{fact:t:wise}) $\ab$ $4$-wise independent uniform r.v.'s $Y_1,\dots,Y_{5\ab}\in [L]$, which they interpret as a random partition $(S_1,\dots, S_L)$ of $[\ab]$ in $L$ parts.
\State Upon observing the sample $\tilde{X}_j=i$
in~\cref{step:goldreich}, player $j$ sends $Y_i$ (corresponding to
  its respective block) represented by  $\numbits$ bits. 
\ForAll{block}\label{step:center_l2:new}
  \State The referee obtains $\ns/N$ independent samples from $(Z_1(\p), \dots, Z_L(\p))$
  \State Knowing the realization of the public coins, it computes the distribution $\tilde{\q}\in\distribs{L}$ corresponding to $(Z_1(\q), \dots, Z_L(\q))$.
      \If{$\normtwo{\tilde{\q}} \leq 2/\sqrt{L}$}\label{step:center_l2:new:test}
          ~it tests if the underlying distribution is $\tilde{\q}$ or $(\gamma/\sqrt{L})$-far from $\tilde{\q}$ in $\lp[2]$, with failure probability $\delta' \gets c/(2+c)$ where $c$ is as in~\cref{theo:identity:z:concentration:anticoncentration:general}. \Comment This uses the test from~\cite{CDVV:14}, stated in~\cref{theo:l2:testing}.
      \Else
          ~it draws a random $\bernoulli{1/2}$ and records it as ``output of the test'' for this block.
      \EndIf
\EndFor 
\State The referee applies the test from~\cref{l:amplify} to the $N$ outputs of the independent tests
(one for each block) and declares the output.
\end{algorithmic}
\caption{A modified, randomness-efficient $\numbits$-bit public-coin protocol for distributed identity testing for
  reference distribution $\q$.}
\label{alg:public-coin-IT:randomness:efficient}
\end{algorithm}
\begin{proof}
The corresponding protocol is provided
in~\cref{alg:public-coin-IT:randomness:efficient}, and it follows the
same structure as~\cref{alg:public-coin-IT}. As discussed
in~\cref{rk:randomness:used}, the two main differences are in~\cref{step:partition:new,step:center_l2:new}. In the
former, we use a random $4$-wise independent partition of $[\ab]$ in
$L$ parts, no longer necessarily equal-sized. This allows us to bring
down the number of public coins to the stated bound, as guaranteed by
the next fact applied with $t=4$: 
\begin{fact}\label{fact:t:wise}
    For any $t\geq 2$, $\ab,\numbits\in\N$, there exists a $t$-wise
    independent probability space $\Omega\subseteq[2^\numbits]^\ab$
    with uniform marginals, and size $\abs{\Omega} = 2^{t(\numbits +
      \clg{\log\ab})}$. Moreover, one can efficiently sample from
    $\Omega$ given $t,\ab,\numbits$. 
\end{fact}
\begin{proof}
The proof relies on a standard construction of $t$-wise independent
$(1/2^\numbits)$-biased random bits via polynomials over an
appropriate finite field. Namely, fixing a field $\mathbb{F}$ of size
$2^{\numbits + \clg{\log\ab}}$ and an equipartition
$F_1,\dots,F_{2^\numbits}$ of $\mathbb{F}$ (so that
$|F_1|=\dots=|F_{2^\numbits}| = 2^{\clg{\log\ab}}$), it suffices to
sample uniformly at random a polynomial $P\in\mathbb{F}_{t-1}[X]$ 
evaluating it at $\ab$ (fixed) points $a_1,\dots,a_\ab\in\mathbb{F}$
yields $t$-wise independent field elements, which correspond to
elements $Y_1,\dots,Y_\ab\in[2^\numbits]$ (where $Y_i =
\sum_{j=1}^{2^\numbits} j\indic{a_i \in F_j}$) with the desired
marginals. 
\end{proof}
In doing so, a new issue arises when applying the {identity}
tester (in $\lp[2]$ distance) of Chan et al.~\cite{CDVV:14} in~\cref{step:center_l2:new}. Note that we can no longer rely on a
centralized uniformity testing algorithm (in $\lp[2]$ distance), as we
did in~\label{alg:public-coin-IT:randomness}. This is because the
resulting reference distribution defined by $(Z_1(\q), \dots,
Z_L(\q))$ is no longer,  in general, the uniform distribution
$\uniformOn{L}$, but some distribution $\tilde{\q}$ on $[L]$. Observe
that this distribution $\tilde{\q}$ is still fully known by the
referee, who is aware of both $\q$ and the realization of the shared
randomness\footnote{Recall that, in contrast to here, the knowledge of shared randomness
  by the referee was not used in~\cref{alg:public-coin-IT}.} (and therefore of $Y_1,\dots,Y_{5\ab}$). 

To handle this issue, we observe that the testing algorithm in
$\lp[2]$ distance of Chan et al. does provide a guarantee beyond
uniformity testing, for the general question of identity testing in
$\lp[2]$ distance. It is, however, a guarantee which degrades with the
$\lp[2]$ norm of the reference distribution (in our case,
$\tilde{\q}$). 
\begin{theorem}[{\cite[Proposition~3.1]{CDVV:14}, with the improvement of~\cite[Lemma II.3]{DK:16}}]\label{theo:l2:testing}
There exists an algorithm which, given distance parameter $\dst>0$, $\ab\in\N$, and $\beta > 0$, satisfies the following. Given $\ns$ samples from each of two unknown distributions $\q,\q'\in\distribs{\ab}$ such that $\beta\geq \min(\normtwo{\q},\normtwo{\q'})$, the algorithm distinguishes between the cases that $\q=\q'$ and $\normtwo{\q-\q'} > \dst$ with probability at least $2/3$, as long as $\ns \gtrsim \beta/\gamma^2$.
\end{theorem}
\noindent We note that the contribution from~\cite[Lemma II.3]{DK:16}
is to explain how to replace the condition $\beta\geq
\max(\normtwo{\q},\normtwo{\q'})$ from~\cite{CDVV:14} by the weaker
$\beta\geq \min(\normtwo{\q},\normtwo{\q'})$. Further, one can as
before amplify the probability of success from $2/3$ to any chosen
constant, at the price of a constant factor in the sample
complexity. We would like to apply this lemma to testing identity to
the $L$-ary distribution $\tilde{\q}$, with distance parameter
$\gamma/\sqrt{L}$ and parameter $\beta \eqdef
\normtwo{\tilde{\q}}$. The desired sample complexity would follow if
we had $\normtwo{\tilde{\q}}\lesssim 1/\sqrt{L}$, since then we would
get 
\[
    \frac{\normtwo{\tilde{\q}}}{(\gamma/\sqrt{L})^2} \lesssim \frac{\sqrt{L}}{\gamma^2}\,.
\]
Of course, we cannot argue that $\normtwo{\tilde{\q}}\lesssim
1/\sqrt{L}$ with probability one over the choice of the random
partition. However, since $F_\q(\q) = \uniformOn{5\ab}$, it is a
simple exercise to check that, over this choice,
\[
\bE{}{\normtwo{\tilde{\q}}^2} = 1/(5\ab) + (5\ab-1)/(5\ab L)  \leq
2/L.
\]
Therefore, letting $c\in(0,1]$ be the constant
  from~\cref{theo:identity:z:concentration:anticoncentration:general},
  we get by Markov's inequality that  $\normtwo{\tilde{\q}}\leq
  2/(\sqrt{cL})$ with probability at least $1-c/2$.  

Since we ran, in~\cref{step:center_l2:new:test}, the identity test with probability of failure $\delta' \eqdef c/(2+c)$, we have the following. When $\p=\q$, each block outputs $1$ with probability at least
\[
 \theta_1 \eqdef \frac{1}{2}\cdot \frac{c}{2} + (1-\delta')(1-\frac{c}{2}) = 1-\frac{c}{4} - (1-\frac{c}{2})\delta' = \frac{c^2 - 2 c + 8}{4 (c + 2)}
\]
while, when $\p$ is $\dst$-far from $\q$, the test for each 
block outputs $0$ with probability greater than
\[
  \theta_2 \eqdef (1-\delta')c = \frac{2c}{c + 2}
\]
so that we have indeed $\theta_1 > 1-\theta_2$. We then conclude the proof as that of~\cref{theo:identity:shared:randomness:ub}, amplifying the probabilities of success by invoking~\cref{l:amplify} and choosing a suitable $N=\Theta(1)$. The total number of public coins used is then at most $N\cdot 4(\clg{\log(5\ab)}+\numbits) = O(\log\ab)$, as claimed.
\end{proof}

%% file: app-identityfromuniformity.tex
\subsection{From uniformity to parameterized identity testing}\label{app:identity:from:uniformity}
In this appendix, we explain how the existence of a distributed
protocol for uniformity testing implies the existence of one for
identity testing with roughly the same parameters, and further even
implies one for identity testing in the \emph{massively parameterized}
sense\footnote{Massively parameterized setting, a terminology borrowed
  from property testing, refers here to the fact that the sample
  complexity depends not only on a single parameter $\ab$ but a
  $\ab$-ary distribution $\q$.} (``instance-optimal'' in the
vocabulary of Valiant and Valiant, who introduced
it~\cite{VV:14}). These two results will be seen as a straightforward
consequence of~\cite{Goldreich:16}, which establishes the former
reduction in the standard non-distributed setting; and
of~\cite{BCG:17}, which implies that massively parameterized identity
testing reduces to ``worst-case'' identity testing. Specifically, we
show the following:

\begin{proposition}
    \label{prop:identity:from:uniformity}
  Suppose that there exists an $\numbits$-bit $(\ab, \dst,
  \delta)$-uniformity testing protocol $\pi$ for
  $\ns(\ab,\numbits,\dst, \delta)$ players.  Then there exists an
  $\numbits$-bit $(\ab, \dst, \delta)$-identity testing protocol
  $\pi'$ against any fixed distribution $\q$ (known to all players),
  for $\ns(5\ab,\numbits,\frac{16}{25}\dst,\delta)$ players.
  
  Furthermore, this reduction preserves the setting of randomness
  (\ie, private-coin protocols are mapped to private-coin protocols).
\end{proposition}
\begin{proof}
  We rely on the result of Goldreich~\cite{Goldreich:16}, which
  describes a mapping $F_{\q}\colon
  \distribs{[\ab]}\to\distribs{[5\ab]}$ such that
  $F_{\q}(\q)=\uniform_{[5\ab]}$ and
  $\totalvardist{F_{\q}(\p)}{\uniform_{[5\ab]}} > \frac{16}{25}\dst$
  for any $\p\in\distribs{[\ab]}$ $\dst$-far from
  $\q$.\footnote{In~\cite{Goldreich:16}, Goldreich exhibits a
    randomized mapping that converts the problem from testing identity
    over domain of size $\ab$ with proximity parameter $\dst$ to
    testing uniformity over a domain of size $\ab'\eqdef\ab/\alpha^2$
    with proximity parameter $\dst'\eqdef (1-\alpha)^2\dst$, for every
    fixed choice of $\alpha\in(0,1)$. This mapping further preserves
    the success probability of the tester. Since the resulting
    uniformity testing problem has sample complexity
    $\bigTheta{\sqrt{\ab'}/{\dst'}^2}$, the blowup factor
    $1/(\alpha(1-\alpha)^4)$ is minimized by $\alpha = 1/5$.} In more
  detail, this mapping proceeds in two stages: the first allows one to
  assume, at essentially no cost, that the reference distribution $\q$
  is ``grained,'' \ie, such that all probabilities $\q(i)$ are a
  multiple of $1/m$ for some $m \lesssim \ab$. Then, the second
  mapping transforms a given $m$-grained distribution to the uniform
  distribution on an alphabet of slightly larger cardinality. The
  resulting $F_\q$ is the composition of these two mappings.
  
  Moreover, a crucial property of $F_\q$ is that, given the knowledge
  of $\q$, a sample from $F_{\q}(\p)$ can be efficiently simulated
  from a sample from $\p$; this implies the proposition.
\end{proof}

\begin{remark}
  The result above crucially assumes that every player has explicit
  knowledge of the reference distribution $\q$ to be tested against,
  as this knowledge is necessary for them to simulate a sample from
  $F_{\q}(\p)$ given their sample from the unknown $\p$. If only the
  referee~$\referee$ is assumed to know $\q$, then the above reduction
  does not go through.
\end{remark}

The previous reduction enables a distributed test for any identity
testing problem using at most, roughly, as many players as that
required for distributed uniformity testing. However, we can expect to
use fewer players for specific distributions. Indeed, in the standard,
non-distributed setting, Valiant and Valiant in~\cite{VV:14}
study a refined analysis termed the \emph{instance-optimal}
setting and showed that the sample complexity of testing identity to
$\q$ is captured roughly by the $2/3$-quasinorm of a 
sub-function of $\q$ obtained as follows: Assuming without loss of generality $\q_1 \geq\q_2\geq \dots \geq \q_\ab \geq 0$, 
let $t\in[\ab]$ be the largest integer that $\sum_{i=t+1}^\ab
q_i \geq \dst$, and let $\q_\dst=(\q_2,\dots,\q_t)$ (\ie, removing
the largest element and the ``tail'' of $\q$). The main result
in~\cite{VV:14} shows that the sample complexity of testing identity
to $\q$ is upper and lower bounded (up to constants) by 
$\max\{\norm{\q_{\dst/16}}_{2/3}/\dst^2,1/\dst\}$ and
$\max\{\norm{\q_\dst}_{2/3}/\dst^2,1/\dst\}$, respectively.

However, it is not clear if the aforementioned reduction of Goldreich
between 
identity and uniformity testing preserves this parameterization
of sample complexity for identity testing. In particular, the
$2/3$-quasinorm characterization does not seem to be amenable to the
same type 
of analysis as that
underlying~\cref{prop:identity:from:uniformity}. Interestingly, a  
different instance-optimal characterization due to Blais, 
Canonne, and Gur~\cite{BCG:17} admits such a reduction,
enabling us to obtain the analogue
of~\cref{prop:identity:from:uniformity} for this massively
parameterized setting.\medskip

To state the result as parameterized by $\q$ (instead of $\ab$), we
will need the definition of a new functional, $\Phi(\q,\gamma)$; 
 see~\cite[Section 6]{BCG:17} for a discussion
on basic properties of $\Phi$ and how it
relates to notions such as the sparsity of $\p$ and the  
functional $\norm{\p_{\gamma}^{-\max}}_{2/3}$ defined in~\cite{VV:14}. For
$a\in\lp[2](\N)$ and $t\in(0,\infty)$, let
\begin{equation}
    \kappa_{a}(t) \eqdef \inf_{a'+a''=a} \left( \normone{a'} +
    t\normtwo{a''} \right)
    \nonumber
\end{equation}
and, for $\q\in\distribs{\N}$ and any $\gamma\in(0,1)$, let
\begin{equation}
  \Phi(\q,\gamma) \eqdef 2\kappa_{\q}^{-1}(1-\gamma)^2\,.
  \nonumber
\end{equation}
It was observed in~\cite{BCG:17} that if $\q$ is supported on at most $\ab$
elements, $\Phi(\q,\gamma) \leq 2\ab$ for all $\gamma\in (0,1)$. Moreover, the sample complexity of testing identity
to $\q$ was shown there to be upper and lower bounded (again up to constants) by 
$\max(\Phi(\q,\dst/9)/\dst^2,1/\dst)$ and
$\Phi(\q,2\dst)/\dst$, respectively. We
are now in a position to state our general reduction.

\begin{proposition}\label{prop:param:identity:from:uniformity}
  Suppose that there exists an $\numbits$-bit $(\ab, \dst, \delta)$-uniformity testing protocol $\pi$ for $\ns(\ab,\numbits,\dst, \delta)$ players.
  Then there exists an $\numbits$-bit $(\ab, \dst, \delta)$-identity
  testing protocol $\pi'$ for any fixed reference distribution $\q$
  (known to all players), 
  for $\ns\mleft( 5\mleft( \Phi(\q,\dst/9)+1 \mright),\numbits, \dst/3, \delta\mright)$ players.
  
  Further, this reduction preserves the setting of randomness (\ie,
  private-coin protocols are mapped to private-coin protocols).
\end{proposition}
\begin{proof}
  This strengthening of~\cref{prop:identity:from:uniformity} stems
   from  the algorithm for identity testing given in~\cite{BCG:17}, which at a
  high-level reduces testing identity to $\q$ of an (unknown) distribution $\p$ to testing identity of $\p|_{S_{\q}(\dst)}$ of $\q|_{S_{\q}(\dst)}$, where $S_{\q}(\dst)$ is the \emph{$(\dst/3)$-effective support}\footnote{Recall
  the \emph{$\dst$-effective support} of a distribution $\q$ is a
  minimal set of elements accounting for at least $1-\dst$ probability
  mass of $\q$.}{} of $\q$; along with checking that $\p$ also only
  puts probability mass roughly $\dst/3$ outside of
  $S_{\q}(\dst)$. The key result of~\cite{BCG:17} relates
  this effective support to the functional $\Phi$ defined above. 
They show (see~\cite[Section 7.2]{BCG:17}) that for all
$\q\in\distribs{\ab}$ and $\dst\in(0,1]$, 
  \begin{equation}\label{eq:bcg:relating:functional}
        \abs{S_{\q}(\dst)} \leq \Phi\bigg(\q,\frac{\dst}{9}\bigg)\,.
  \end{equation}
See~\cref{fig:distributions} for an illustration.
\begin{figure}[ht]\centering
  \begin{tikzpicture}[x=1pt,
  y=10pt] \pgfmathsetmacro{\xmax}{300} \pgfmathsetmacro{\ymax}{10} %
  Grid %
  (\xmax,\ymax); \draw [<->] (0,\ymax) node[above] {$\q(i),\p(i)$} --
  (0,0) -- (\xmax,0) node[right] {$i$};

  \pgfmathsetseed{897273} %

  \pgfmathsetmacro{\rbuckets}{40} \pgfmathsetmacro{\buckwidth}{(\xmax/\rbuckets)} \pgfmathsetmacro{\T}{\rbuckets*0.75} \pgfmathsetmacro{\R}{\T+1}

      \node [below] at (\xmax,-0.25) {$\ab$}; \node [below] at
      (0,-0.25) {$1$}; \node [below] at ({\T*\buckwidth},-0.25)
      {$\ab_\dst$}; \node [below] at ({\T*\buckwidth/2},-0.25)
      {$S_\q(\dst)$};

    \draw[thin, dotted] 
    \foreach \i in
    {1,...,\rbuckets}{ ({\i*\buckwidth},0) -- ({\i*\buckwidth},\ymax)
    }; \draw ({\T*\buckwidth},0) -- ({\T*\buckwidth},\ymax);

        \foreach \i in{1,...,\rbuckets}{
        \pgfmathsetmacro{\cointoss}{random}
        \pgfmathsetmacro{\cointoss}{(random()-0.5)/10+(random()-1/2)*abs(\cointoss)/\cointoss};
        \draw[thin,blue]      ({(\i-1)*\buckwidth},{\ymax-1.25*sqrt(\i)}) --
        ({(\i)*\buckwidth},{\ymax-1.25*sqrt(\i)}); \draw[thin,red]
        ({(\i-1)*\buckwidth},{\ymax-1.25*sqrt(\i)+\cointoss}) --
        ({(\i)*\buckwidth},{\ymax-1.25*sqrt(\i)+\cointoss});
        };
        \foreach \i in {\R,...,\rbuckets}{ 
        \fill[thick,blue,opacity=0.15] ({(\i-1)*\buckwidth},0) --
        ({(\i-1)*\buckwidth},{\ymax-1.25*sqrt(\i)}) --
        ({(\i)*\buckwidth},{\ymax-1.25*sqrt(\i)}) --
        ({(\i)*\buckwidth},0) -- cycle;          
        };
        \node [] at ({(\T*\buckwidth+\xmax)/2},{(\ymax -
        1.25*sqrt((\T+\xmax/\buckwidth)/2))/2})
        {$\dst$};
        \end{tikzpicture}\caption{\label{fig:distributions}The
        reference distribution $\q$ (in blue; assumed non-increasing
        without loss of generality) and the unknown distribution $\p$
        (in red). By the  reduction above, testing equality of $\p$ to
        $\q$ is tantamount to
        (i)~determining $S_\q(\dst)$, which depends only on $\q$;
        (ii)~testing identity for the conditional distributions of $\p$ and
        $\q$ given $S_\q(\dst)$, and (iii)~testing that $\p$
        assigns at most $O(\dst)$ probability to the complement of $S_\q(\dst)$.}
\end{figure}
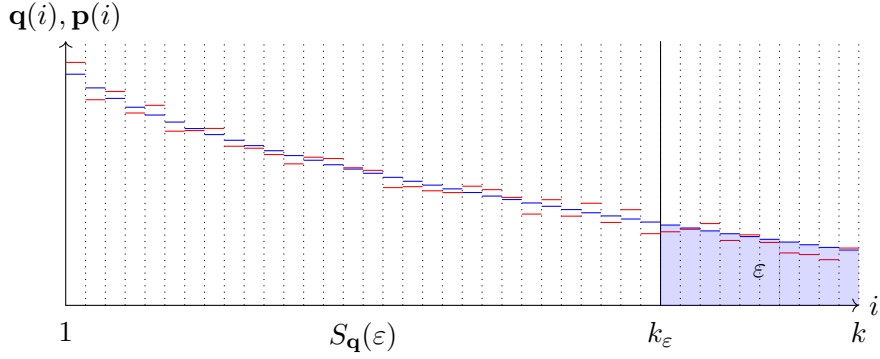

\noindent The protocol $\pi'$ then works as follows:
\begin{enumerate}
  \item Given their knowledge of $\q$ and $\dst$, all players (and the referee) compute $S \eqdef S_{\q}(\dst)$. Consider the following mapping $G_\q\colon\distribs{[\ab]}\to  \distribs{S\cup\{\bot\}}$. For any $\p'\in \distribs{[\ab]}$,
  \[
  G_\q(\p')(x) =
        \begin{cases}
            \p'(x), &\text{ if } x\in S,\\
            \p'([\ab]\setminus[S]), &\text{ if } x = \bot.
        \end{cases}
  \]
  Note that all players have full knowledge of $\tilde{\q} \eqdef G_\q(\q)$. Further, each player, given their sample from the (unknown) $\p$, can straightforwardly obtain a sample from $\tilde{\p} \eqdef G_\q(\p)$.
  \item All players (and the referee) compute $\ab'\eqdef 5(|S|+1)$, and the mapping $F_{\tilde{\q}}\colon \distribs{S\cup\{\bot\}}\to  \distribs{\ab'}$ (as in the proof of~\cref{prop:identity:from:uniformity}). From properties of $F_\q$ described in the proof of~\cref{prop:identity:from:uniformity},  $F_{\tilde{\q}}(\tilde{\q}) = \uniformOn{\ab'}$.
  \item Each player converts their sample from the (unknown) distribution $\tilde{\p}$ into a sample from the (unknown) distribution $F_{\tilde{\q}}(\tilde{\p})$. (Recall that this is possible given the knowledge of $\tilde{\q}$, as stated in the proof of~\cref{prop:identity:from:uniformity}.)
  \item The players and the referee execute the purported
    $\numbits$-bit uniformity testing protocol $\pi$ on their samples
    from $F_{\tilde{\q}}(\tilde{\p})$, with parameters $(\ab', \dst/3,
    \delta)$. The output of $\pi'$ is then that of $\pi$. 
\end{enumerate}

If $\p=\q$, then $\tilde{\p}=\tilde{\q}$ and thus $F_{\tilde{\q}}(\tilde{\p}) = F_{\tilde{\q}}(\tilde{\q}) = \uniformOn{\ab'}$, so that the protocol $\pi$ returns 1 with probability at least $1-\delta$. On the other hand, if $\totalvardist{\p}{\q} > \dst$, then
\begin{align*}
    2\totalvardist{\tilde{\p}}{\tilde{\q}} 
    &= \sum_{x\in S} \abs{\p(x)-\q(x)} + \abs{\p(\bar{S})-\q(\bar{S})}
    = 2\totalvardist{\p}{\q} - \sum_{x\in \bar{S}} \abs{\p(x)-\q(x)} + \abs{\p(\bar{S})-\q(\bar{S})} \\
    &\geq 2\totalvardist{\p}{\q} - (\p(\bar{S})+\q(\bar{S})) + \abs{\p(\bar{S})-\q(\bar{S})} 
    = 2\totalvardist{\p}{\q} - 2\min(\p(\bar{S}),\q(\bar{S})) \\
    &> 2\dst - 2\cdot \frac{\dst}{3} = \frac{4}{3}\dst
\end{align*}
\ie, $\totalvardist{\tilde{\p}}{\tilde{\q}} > 2\dst/3$. Recalling the guarantee of Goldreich's reduction (as described in the proof of~\cref{prop:identity:from:uniformity}), this in turns implies that
$
    \totalvardist{ F_{\tilde{\q}}(\tilde{\p}) }{ \uniformOn{\ab'} } \geq (16/25) \cdot 2\dst/3 > \dst/3
$, 
and therefore the protocol $\pi$ must return 0 with probability at least $1-\delta$.

\noindent To conclude, in view of~\eqref{eq:bcg:relating:functional},
the number of players required by $\pi'$ is
\[
    \ns(\ab',\numbits, \dst/3, \delta) 
    = \ns(5(\abs{S_{\q}(\dst)}+1),\numbits, \dst/3, \delta)
    \leq \ns\mleft( 5\mleft( \Phi(\q,\dst/9)+1 \mright),\numbits, \dst/3, \delta\mright)\,,
\]
as claimed.  
\end{proof}